\documentstyle[12pt,up_thesis]{report}

\begin{document}
\title{Studies of fractal structures and processes using methods of the 
fractional calculus}
\author{Kiran M. Kolwankar}

\dept{ Physics }
\guide{Dr. Anil D. Gangal}
\submitdate{September 1997}

\coguidefalse
\tablespagefalse

\beforepreface
\prefacesection{Acknowledgement}I take this opportunity, on submission of my thesis, 
to express my gratitude to those all who helped
me, in various capacities, during the work.

I am extremely indebted to Dr. A. D. Gangal for his valuable guidance
but for which this thesis would never have seen daylight.
His continual insistence on quality and perfection has certainly
influenced my understanding and approach to my work.
He was always forthcoming with all the helps whenever I needed them.

I have benefited a great deal through illuminating discussions,
academic as well as nonacademic, with Dr. R. E. Amritkar and Dr. N. M. Gupte. 
Dr. H. Bhate has, time and again, helped me out with various technical difficulties.
My work has also been influenced by discussions with Dr. Athavale,
Dr. U. V. Naik-Nimbalkar, Prof. P. V. Panat and Prof. Kupsch.
I thank Prof. Kanhere and Prof. D. S. Joag for making computer
and assorted facilities 
available to me.

Prof. P. L. Kanitkar, Head, Department of Physics, deserves very special
thanks for generously allowing me to use all kinds of infrastructural 
facilities. I thank all the non--teaching staff for their kind help.

I cherish my long and intimate association with friends who have made my
stay in Pune real fun whenever the going was difficult.
Discussions with Manojit have always led me to  better
understanding. I also thank Prashant, Ashutosh and Anil for many helpful discussions.
Many thanks to Sagar for help with the computers from time to time.
I would also like to thank Jolly, Gauri, Nandini, Trushant, Nimish, Raja,
Tunna, Lanke, Ajay, Ajita, Sahu, Bala,
Pradeep, Abhijat, Uday, Gore and all others 
for helping me during various stages of my work.
I thank Manojit again for correcting my English whenever I wrote, 
including this page.

I thank my family members for constant support and encouragement.

Last but not the least, I thank CSIR for financial assistance.

\prefacesection{Abbreviations}{\bf List of commonly used abbreviations}

LFD -- Local fractional derivative

LFDE -- Local fractional differential equation

FPK -- Fokker-Planck-Kolmogorov

FFPK -- Fractional Fokker-Planck-Kolmogorov

\afterpreface
\prefacesection{Abstract}
Fractional calculus \cite{OS, MR} is a branch of classical mathematics, 
which deals with
the generalization of operations of differentiation and integration to 
fractional order. Such a generalization is not merely a mathematical curiosity
but has found applications in various fields of physical sciences. Various
 differential equations of physics, such as wave equations, diffusion equations
\cite{GR,SW,Jum}, etc, have been generalized to include fractional orders. 
Recently there has
been a surge of activity involving the use of fractional calculus 
in studying phenomena relating to fractal structures and processes.

Fractals \cite{Man,Fal} are sets having fractional dimension. 
They appear naturally
in nonlinear and nonequilibrium phenomena in different forms and context. 
Frequently attractors of chaotic dynamical systems are fractals 
(e.g. fractal dust or nowhere
differentiable fractal curve). 
We note that the graph of a nowhere differentiable functions is generally a
fractal set.
Also typical paths of a Brownian particle or 
Feynman paths in Quantum Mechanics \cite{FH,AW}
are nowhere differentiable curves.
There are processes, e.g. L\'evy process, which generate a fractal set.
Processes which have evolution taking place at time instances lying on
fractal sets can be used to model various transport phenomena, e.g., transport
in spongy or disordered materials, transport in chaotic Hamiltonian systems.

Ordinary calculus
is inadequate to characterize and handle such irregular curves and surfaces.
Also we felt a need to develop calculus which will describe phenomena 
(such as diffusion) on 
fractals. It is an interesting question to ask: whether the formalism of
fractional calculus can provide such a framework. The aim of the thesis
is to explore this possibility. And our answer is in the affirmative.

The first chapter of the thesis contains motivation for the work that
we have carried out.
It also gives the overview and defines the scope of the thesis.

The second chapter gives the introduction and brief review
of the related fields. Which includes a
survey of the relevant part of the fractional calculus.
Definitions and examples of fractals are also given.
A classical notion of fractional differentiability has also been briefly reviewed.

The third chapter starts with the definition of the local fractional 
derivative(LFD), for order $q$, $0<q<1$, along with its 
motivation.
The LFD of order $q$ of a function $f(y)$ is defined by
\begin{eqnarray}
I\!\!D^qf(y) =  {\lim_{x\rightarrow y}}
{{d^q[f(x)-f(y)]}
\over{[d(x-y)]^q}} \;\;\;0 < q \leq 1\label{deflocginabs}
\end{eqnarray}
where the derivative  on the RHS is the
 Riemann-Liouville fractional derivative \cite{OS,MR}, viz., for $0<q<1$
\begin{eqnarray}
{{d^qf(x)}\over{[d(x-a)]^q}}={1\over\Gamma(1-q)}{d\over{dx}}{\int_a^x{{f(y)}
\over{(x-y)^{q}}}}dy.\label{def2}
\end{eqnarray}
This definition is then generalized to include all positive orders
$q$. Further we have also generalized this definition for functions of
several variables. All the definitions have been illustrated with the
help of examples.  After this introduction of the definitions
a fractional Taylor expansion has been derived which involves these LFDs.
Various consequences of this local fractional Taylor expansion have been
discussed.

The fourth chapter deals with the question of fractional differentiability
properties of continuous but nowhere differentiable functions. 
It starts with the introduction to nowhere differentiable functions. 
Then a specific historically important
example of Weierstrass function is considered. Using the definition of LFD
introduced in the last chapter we have 
proved the striking result that this function is locally
fractionally differentiable upto the critical order $2-s$ where $s$ is the box
dimension of the graph of the function. 
This is followed by some general results 
which establish a connection between the critical order and the 
local H\"older exponent of the function.

Use of LFD in studying regularity properties of the irregular functions
has been considered in the fifth chapter. The chapter begins with the
introduction and examples of multifractal functions. 
This follows by the demonstration, with the
help of examples, of the fact that LFD can be used to detect isolated 
singularities
as well as singularities which are masked by stronger ones.
Further it has been shown that the LFD can be used even in the case
where singularities are not isolated and local H\"older exponent is not same 
at all points. We have used a specific example of a selfsimilar multifractal 
function constructed by Jaffard for this purpose.

The sixth chapter introduces a new kind of equation called local fractional
differential equation (LFDE) which is an equation involving LFDs. This
is a generalization of usual differential equations to fractional order
keeping the local nature intact. A simple example of such an equation, viz.,
\begin{eqnarray}
I\!\!D^q_xf(x) = g(x), \label{eq:slfdeinabs}
\end{eqnarray}
is considered. It is found that the solution to this equation does not exist
when $g(x)=\mbox{Const.}$. But it admits a finite and nontrivial solution
if $g(x)$ has fractal support. It is argued that these
LFDEs are suitable to describe phenomenon taking place in fractal space and 
time.
As one possible application of LFDE to physical situation 
a local fractional 
Fokker-Planck equation has been derived starting 
from first principles. A specific case of this equation
is solved and found to give rise to a subdiffusive solution.

The seventh and final chapter concludes the thesis. Along with the
concluding remarks we discuss the further question which have come up.

\newtheorem{lemma}{Lemma}
\newtheorem{thm}{Theorem}
\newtheorem{cor}{Corollary}
\newtheorem{defn}{Definition}
\chapter{Motivation and Overview} \label{ch:0}

\begin{quote}
{\it "One might encounter instances where using a function without derivative
would be simpler than using one that can be differentiated. When this happens,
the mathematical study of irregular continua will prove its practical value."}

\rightline{\ldots{ Perrin}}
\end{quote}

Till the second half of the last century, it was believed that a continuous
function must be differentiable at least at some point. 
This view changed in 1872, when
Weierstrass constructed a function which was continuous but nowhere differentiable. 
Subsequently, a host of instances of other such functions was developed
(historical comments are summarized in~\cite{0Kor}).
For about three decades, these functions were considered as pathological
cases and were used only as counter examples. 
As can be seen from the above quotation (see~\cite{0Man}), it was Perrin who
 recognized their relevance to the physical world.
In his study on Brownian motion, he realized that the path of a
Brownian particle can be modeled by such irregular functions.

Subsequent advances in nonlinear and nonequlibrium phenomena revealed many
more examples of the occurrence of such functions in physical theories.
These include, among others, various generalizations of the Brownian
motion~\cite{0Man,0Fal,0MV}
(e.g., fractional Brownian motion), Feynman paths in quantum mechanics,
attractors of some dynamical 
systems~\cite{0KMY,0FOY}, velocity field of
turbulent fluid~\cite{0CPS,0CP}, etc. 
Before proceeding to the discussions of these irregular functions
we would like to point out the developments in the
mathematical theory of sets of fractional dimension showed that 
generally the graphs of such functions are fractals 
(hence we may alternatively call these functions as fractal functions).

\begin{itemize}
\item Jean Perrin (who won the Nobel Prize for his work on Brownian motion)
pointed out that the path
followed by a Brownian particle is irregular on all length scales
and hence, one can not draw a tangent to this path. It was shown 
later~\cite{0Man} 
that the path followed by the Brownian particle in $I\!\!R^E$ with
embedding dimension $E\geq 2$ is a fractal with
dimension 2. Another similar example comes from quantum mechanics
where typical Feynman paths  are  differentiable nowhere. 
Observed path of a particle in quantum mechanics
was shown~\cite{0AW} to be a fractal curve with Hausdorff dimension 2.
The work on Brownian motion (see~\cite{0Fal}) showed that the graphs of 
components of Brownian paths are nowhere differentiable and have
dimension $3/2$. Brownian motion was later generalized
~\cite{0Man,0Fal,0MV} to 
{\it fractional Brownian motion}. This process is known to give rise to
graphs which have dimensions between 1 and 2.

\item Turbulent fluid system is another source of irregular functions.
In these systems, the inherent nonlinear nature 
in the turbulent regime is believed to give rise to the 
irregularity. Passive scalars advected by a turbulent fluid have been
shown~\cite{0CPS,0CP} to have isoscalar surfaces which are highly
irregular, in the limit of diffusion constant going to zero.
Velocity fields of a turbulent fluid at low viscosity~\cite{0FP}
have been shown to be multifractal.

\item In the study of dynamical systems theory, attractors of some systems
are found to be continuous but nowhere differentiable~\cite{0KMY}.
We will consider one specific example~\cite{0KMY,0FOY}
of the dynamical system which gives 
rise to such an attractor. Consider the following map.
\begin{eqnarray}
x_{n+1}&=&2x_n+y_n\;\;\;\;\mbox{mod}\;1, \nonumber\\
y_{n+1}&=&x_n+y_n\;\;\;\;\;\;\mbox{mod}\;1,\\
z_{n+1}&=&\lambda z_n + \cos(2\pi x_n) \nonumber
\end{eqnarray}
where $x$ and $y$ are taken mod 1 and $z$ can be any real number.
In order to keep $z$ bounded, $\lambda$ is chosen between 0 and 1.
The equations for $x$ and $y$ are independent of $z$.  The $x-y$
dynamics are chaotic and are unaffected by the value of $z$.
It was shown in~\cite{0KMY} that if $\lambda > 2/(3+\sqrt{5})$,
then the attractor of this map is a nowhere differentiable torus
(i.e., it has same topological form as a torus but is nowhere differentiable). 
A simpler
example of a map whose repeller is a graph of a nowhere differentiable function 
is considered in \ref{se:1ch3}.

\item A variety of natural and industrial processes lead to formation
of rough surfaces. Crystal growth, vapor deposition, grinding, erosion, 
fracture are a few examples of processes leading to 
such surfaces. To understand the dynamics
of growth and formation of rough interfaces is one of the challenging
problems. Formalism based on concepts of fractals
~\cite{0Fam,0Vic,0Fed} has become a standard
tool in the study of such surfaces.

\item Many problems of physical interest involve solving a partial differential
equation with Dirichlet boundary conditions. A variety of
methods exist to solve such equations with smooth boundaries. But solving
such equations with nowhere differentiable surfaces as a boundary presents a
great difficulty. Such problems arise when one wants, for instance, to find
field near a charged rough surface~\cite{0SM} or to study vibrations of a
drum with fractal boundary~\cite{0SG}.

\end{itemize}

The world is full of such nonlinear and nonequilibrium phenomena. However, 
the complicated nature
of these phenomena make them analytically intractable. One of the reasons
for the lack of availability of analytical methods can be traced to 
almost ubiquitous appearance
of fractals in these phenomena. 
It is clear from the examples given above that the graph of the fractal
functions can arise as attractors or repellers of dynamical systems,
as paths of particles performing random motion, etc. 
Moreover, general fractal sets appear at numerous places in 
nonlinear theories. 
There are many interesting problems in which one has to deal with the
phenomena taking place in fractal space or time. For instance,
one may consider the diffusion on fractals, where underlying
space is a fractal. Such a model is used to describe
diffusion in disordered systems. Diffusion in which transitions
take place at time instances which lie on a fractal set, can be used
to model diffusion in presence of traps. The usual differential calculus,
which is used to describe phenomena in Euclidean space and time, is inadequate
to handle these cases.

It is interesting to investigate whether
{\it fractional calculus}, which generalizes the operation of derivation and
integration to fractional order, can provide a possible calculus to deal with 
fractals.
In fact there has been a surge of activity in recent times which supports
this point of view. This possible connection between fractals and
fractional calculus gives rise to various interesting questions.
For example, one may ask that since an $n$-fold integral gives
an $n$-dimensional volume, whether a suitably defined `$\alpha$-fold'
integral ($\alpha$ noninteger) 
would give rise to $\alpha$-dimensional measure. Or, how would one generalize
a formula for finding length of a smooth curve to one which finds  fractal measure of a  continuous but nowhere differentiable curve?
These lead to a more specific problem of studying {\it fractional 
differentiability}
properties of nowhere differentiable functions, and, the possible
relation of the order of differentiability with 
the dimension of the graph of the function.

The aim of the thesis was to investigate the issues discussed above.
The strategy we employed was as follows. We  first concentrated on the fractal 
(irregular and possibly nowhere
differentiable) functions and study their fractional differentiability
properties. Our attention was focused on the local scaling
property of these functions.
This led us to construct a quantity called {\it local fractional derivative}.
It was then used to study pointwise behavior of irregular functions. 
Finally, we demonstrated that these local fractional derivatives can
be used to construct equations to deal with phenomena involving fractals.
It can also be seen as a first step towards
development of much sought after calculus for describing phenomena 
taking place in 
fractal space and time.

Multifractal measure has been the object of many investigations
\cite{0FP,0BPPV,0HJKPS,0CLP,0JKP,0Man1,0MS,0CS}, with a host of
interesting applications. Its importance also stems
from the fact that such measures are natural measures to be used in the
analyses of many physical phenomena~\cite{0Vic,0Fed}. 
Rigorous mathematical results concerning multifractals were
developed in~\cite{0CM}.
However, it may so happen that the object
one wants to understand is a function (e.g. a fractal or multifractal signal),
rather than a set or a measure. For instance, one would like to
characterize the velocity field of fully developed turbulence.
Studies on fluid turbulence have shown the existence of multifractality  
 in velocity fields of a turbulent fluid at low viscosity~\cite{0FP}.

In the light of these needs there has been a renewed interest, 
chiefly among mathematicians, in studying
pointwise behavior  of multifractal functions.
New tools have been developed to study {\it local} behavior of functions,
which will be of practical use.
Notable among them is the use of Wavelet transforms
~\cite{0Hol,0Jaf1,0ABM,0JM} which
 has some success in this effort. 
In this 
thesis we develop and use local fractional derivatives
to study pointwise behavior of multifractal functions and point out
some advantages over the existing methods.

The further plan of the thesis is to begin by surveying some relevant
fields in the next chapter.
With this introduction, we proceed in the thesis with the characterization
of the local scaling property of the fractal functions. In order to do this,
we modify the definition of the usual fractional derivative and
introduce the notion of local fractional derivative
(chapter~\ref{ch:2}). It is shown, in the same chapter, that 
these local fractional derivatives
appear naturally in a fractional Taylor expansion. In chapter~\ref{ch:3}, we
study the local fractional differentiability properties of the nowhere
differentiable functions. Choosing the Weierstrass function as a prototype
example, we show that this function is locally fractionally differentiable 
at every point, upto some critical order which lies between 0 and 1. This 
critical order is related to the fractal dimension of the graph of the
function. We then demonstrate (chapter~\ref{ch:4}) that the local fractional derivatives can be used to study pointwise behavior of the multifractal 
functions.
In chapter~\ref{ch:5}, we introduce the local fractional differential equations.
These equations are expected to describe phenomena in fractal space-time.
The algorithm for the solutions of special cases of
these equations has also been discussed in this chapter.
As an application, we derive the fractional analog of the Fokker-Planck
equation. A special case of this equation is then solved and shown to have a
subdiffusive solution. The thesis ends with  the concluding remarks.

\chapter{Introduction} \label{ch:1}

The purpose of this chapter is to present a brief introduction to some of the
background material
needed for the theme of the thesis. Since it deals with applications of
fractional calculus to fractals, introduction to both these fields is
essential. In the first section we consider the fractional
calculus, giving various definitions of fractional derivatives and integrals,
their properties and few examples. In the next section we deal with
fractals. We list desirable properties of the definition of dimension and,
in particular,
discuss two definitions in detail. We also discuss some applications of
fractals in the physical systems. In a separate section we review 
earlier applications of fractional calculus to scaling phenomena.
The last section of this chapter is devoted to a brief review of classical
notions of fractional differentiability.

\newpage
\section{Fractional Calculus} \label{se:1ch1}

\begin{quote}
{\it "Thus it follows that $d^{1\over 2}x$ will be equal to $x\sqrt{dx:x}$, an
apparent paradox, from which one day useful consequences will be drawn."}

\rightline{\ldots Leibniz}
\end{quote}

Fractional calculus~\cite{1OS,1MR,1Ros} 
is a branch of classical mathematics which deals with
generalization of the operations of derivation and integration to fractional
order (even to irrational and complex orders - in this sense it is a misnomer). Though little known, the subject is by no means new. Historically
the interest in the subject dates all the way back to the year 1695,
 when Leibniz made the
above prophetic remark in a letter to L'Hospital, quoted in~\cite{1OS}
(this was the era during which Newton and Leibniz independently developed
the ideas of classical calculus).
During the next two centuries the subject slowly
matured through the works of the mathematicians
Euler, Lagrange, Laplace, Abel, Liouville, Riemann, to name but a few.

\subsection{Definitions}
There exists a multitude of definitions of fractional derivatives and integrals,
which have different origins and are not necessarily equivalent. We shall list few of 
them here. Let's start with the definition given by Grunwald (see~\cite{1OS}),
which directly extends and unifies the notions of quotient of differences
and Riemann sums. According to
this definition a differintegral of order $q$ of a function 
$f:I\!\!R \rightarrow I\!\!R$ is given by
\begin{eqnarray}
{d^qf \over{[d(x-a)]^q}} = {\lim_{N\rightarrow \infty}
\left\{{{\left[{x-a\over N}\right]^{-q}}\over \Gamma(-q)} \sum_{j=0}^{N-1}
{\Gamma(j-q)\over\Gamma(j+1)} f(x-j\left[{x-a\over N}\right])\right\}},
\end{eqnarray}
where $q$, $a \in I\!\!R$ and $\Gamma(p)$ is the usual gamma function
defined by the integral
\begin{eqnarray}
\Gamma(p) = \int_0^\infty x^{p-1} e^{-x} dx, \;\;\;\; p>0,
\end{eqnarray}
and by the relation $\Gamma(p+1)=p\Gamma(p)$. For $q>0$ the above formula yields
derivatives of order $q$, and for $q<0$ it gives integrals of order $q$.
We shall follow this convention throughout.  It may be noted that
{\it the fractional derivative depends on the lower limit `$a$'}. 
The above definition involves the
fewest restrictions on the functions to which it applies and,
unlike some other definitions introduced below, does not 
use ordinary derivatives and integrals of the function. However,
in practice it is somewhat difficult
to use this definition, except in the cases of simple functions. 

The most frequently used definition of a fractional integral
is the Riemann--Liouville definition. According to this definition, a
fractional integral of order $q$ of a function $f$ is given by
\begin{eqnarray}
{{d^qf(x)}\over{[d(x-a)]^q}}={1\over\Gamma(-q)}{\int_a^x{{f(y)}\over{(x-y)^{q+1}}}}dy
\;\;\;{\rm for}\;\;\;q<0,\label{defRLi}
\end{eqnarray}
where the lower limit $a$ is some real number. The fractional derivative
of order $q$ is 
\begin{eqnarray}
{{d^qf(x)}\over{[d(x-a)]^q}}&=&{1\over\Gamma(n-q)}{d^n\over{dx^n}}
{\int_a^x{{f(y)}\over{(x-y)^{q-n+1}}}}dy
\;\;\;{\rm for}\;\;\; n-1<q<n\label{defRLd} \\
&=& \sum_{k=0}^{n-1}{(x-a)^{-q+k}f^{(k)}(a) \over \Gamma(-q+k+1)} +
{{d^{q-n}f^{(n)}}\over{[d(x-a)]^{q-n}}} \;\;{\rm for}\;\; 
n-1<q<n.\label{defRLd2}
\end{eqnarray}
As is clear, this definition uses the concepts of ordinary derivatives and integration.
Since it amounts to evaluating an integral, it is more 
convenient to use.
It can be shown~\cite{1OS} that
the Grunwald definition and the Riemann-Liouville definition are equivalent
to each other.

Hermann Weyl defined fractional integral of order $q$ of a function $f$ by
\begin{eqnarray}
{{d^qf(x)}\over{[dx^q]}}={1\over\Gamma(-q)}{\int_{-\infty}^x
{{f(y)}\over{(x-y)^{q+1}}}}dy
\;\;\;{\rm for}\;\;\;q<0.\label{defWi}
\end{eqnarray}
This definition is suitable for periodic functions as
it leaves the periodicity of the function unaffected, unlike in the cases
of previous two definitions.
 Because of this property, Zygmund~\cite{1Zyg} 
has used this definition extensively in working with trigonometric series.
Many more definitions of fractional derivatives and integrals can be found
in references~\cite{1OS,1MR}.

\subsection{Properties}
In this subsection we review some of the properties of the operations 
of differentiation and integration to arbitrary order~\cite{1OS,1MR,1Ros}.

\subsubsection{Linearity and homogeneity}
From the definitions introduced above it is clear that the
differintegral satisfies 
\begin{eqnarray}
{{d^q[f_1+f_2]}\over{[d(x-a)]^q}} = {{d^qf_1}\over{[d(x-a)]^q}} + 
{{d^qf_2}\over{[d(x-a)]^q}}
\end{eqnarray}
and
\begin{eqnarray}
{{d^q[Cf]}\over{[d(x-a)]^q}} = C{{d^qf}\over{[d(x-a)]^q}}.
\end{eqnarray}

\subsubsection{Change of scale}
When the argument of the function is scaled by a factor $\beta$, 
the differintegral
satisfies 
\begin{eqnarray}
{{d^qf(\beta x)}\over{[dx]^q}} = \beta^q {{d^qf(\beta x)}\over{[d(\beta x)]^q}}.
\end{eqnarray}
For a more general formula, with nonzero lower limit $a$, see~\cite{1OS}.

\subsubsection{Leibniz's rule}
Osler~\cite{1Os1} proved a general result for a differintegral of a product of
two functions. It is given by
\begin{eqnarray}
{{d^q[fg]}\over{dx^q}} = \sum_{j=-\infty}^{\infty}
{\Gamma(q+1)\over{\Gamma(q-\gamma-j+1) \Gamma(\gamma+j+1)}}
{{d^{q-\gamma-j}f}\over{dx^{q-\gamma-j}}}{{d^{\gamma+j}g}\over{dx^{\gamma+j}}},
\end{eqnarray}
where $\gamma$ is arbitrary. He used~\cite{1Os1,1Os2} this result to generate certain infinite
series expansions which interrelate special functions of mathematical physics.

\subsubsection{Chain rule}
If $\phi$ is an analytic function and $f$ a sufficiently differentiable 
function, then the chain rule for fractional differentiation is given 
by~\cite{1OS}
\begin{eqnarray}
{{d^q\phi(f(x))}\over{[d(x-a)]^q}}\!=\!{[x-a]^{-q} \over \Gamma(1-q)} \phi(f(x))
\!+\!\! \sum_{j=1}^\infty qC_j {[x-a]^{j-q}j! \over \Gamma(j-q+1)} 
\!\sum_{m=1}^j \phi^{(m)} \!\sum \! \prod_{k=1}^j \! {1\over P_k!}
\big[{f^{(k)}\over k! }\big]^{P_k}\!,
\end{eqnarray}
where $\sum$ extends over all combinations of nonnegative integer values
of $P_1, P_2, \ldots, P_n$ such that $\sum_{k=1}^n k P_k = n$ and 
$\sum_{k=1}^n  P_k = m$. 
This is a generalization of Fa\'a de Bruno's chain rule for integer order
derivatives.
The complicated nature of this formula makes it
less useful.

\subsubsection{Differintegration term by term}
Rules for distributing fractional derivative and integral operators over
an infinite series are the same as those for ordinary derivatives and
integrals. Let us first consider the case of $q\leq 0$. Suppose the
infinite series of functions $\sum f_j$ converges uniformly in
$0< |x-a| < X$, then~\cite{1OS}
\begin{eqnarray}
{{d^q}\over{[d(x-a)]^q}} \sum_{j=0}^\infty f_j = \sum_{j=0}^\infty
{{d^qf_j}\over{[d(x-a)]^q}},\;\;\;\; q\leq 0,
\end{eqnarray}
and the right-hand side also converges uniformly in $0< |x-a| < X$.
For $q>0$, if the infinite series $\sum f_j$ as well as the series
$\sum {{d^qf_j}/{[d(x-a)]^q}} $ converges uniformly in $0< |x-a| < X$, then
\begin{eqnarray}
{{d^q}\over{[d(x-a)]^q}} \sum_{j=0}^\infty f_j = \sum_{j=0}^\infty
{{d^qf_j}\over{[d(x-a)]^q}},\;\;\;\; q > 0,
\end{eqnarray}
for $0< |x-a| < X$.

\subsubsection{Fractional Taylor series}
In~\cite{1Osl} Osler generalized the Taylor series to include
fractional derivatives. He proved a very general result. We consider
a special case which is called the Taylor-Riemann series:
\begin{eqnarray}
f(z) = \sum_{n=-\infty}^{\infty} {(z-z_0)^{n+\gamma} \over \Gamma(n+\gamma+1)}
{d^{n+\gamma}f(z) \over [d(z-b)]^{n+\gamma}} \vert_{z=z_0},
\end{eqnarray}
where $b \neq z_0$ and $f(z)$ is analytic. Osler used this generalization
to study some special functions of mathematical physics. Please note that
this series contains terms with negative powers of $(z-z_0)$ and hence
it is not suitable for use in standard approximating schemes.

\subsubsection{Composition rule}
Here we study~\cite{1OS} the relationship between two operators
\begin{eqnarray}
{{d^q}\over{[d(x-a)]^q}}{{d^Qf}\over{[d(x-a)]^Q}} \;\;\; \mbox{and} \;\;\;
{{d^{q+Q}f}\over{[d(x-a)]^{q+Q}}}.
\end{eqnarray}
When $f=0$,
\begin{eqnarray}
{{d^q}\over{[d(x-a)]^q}}{{d^Q[0]}\over{[d(x-a)]^Q}} =
{{d^{q+Q}[0]}\over{[d(x-a)]^{q+Q}}} = 0
\end{eqnarray}
and the composition rule is trivially satisfied. When $f\neq 0$, the
composition rule is satisfied if
\begin{eqnarray}
f-{{d^{-Q}}\over{[d(x-a)]^{-Q}}}{{d^Qf}\over{[d(x-a)]^Q}} = 0
\end{eqnarray}
and if this condition is not satisfied then we have
\begin{eqnarray}
{{d^q}\over{[d(x-a)]^q}}{{d^Qf}\over{[d(x-a)]^Q}} &=&
{{d^{q+Q}f}\over{[d(x-a)]^{q+Q}}} \nonumber\\
&& - {{d^{q+Q}}\over{[d(x-a)]^{q+Q}}}
\{ f-{{d^{-Q}}\over{[d(x-a)]^{-Q}}}{{d^Qf}\over{[d(x-a)]^Q}}\}.
\end{eqnarray}
It can be shown that when $Q<0$, the composition rule is satisfied for an arbitrary
differintegrable function $f$. In fact if $f(x)/(x-a)^p$ is finite at $x=a$,
then the composition rule is valid for $Q<p+1$.

\subsection{Examples}
In general, finding fractional derivative or integral
of a function can turn out to be very complicated. In this section we
consider few examples of familiar functions for which
the job is relatively easy~\cite{1OS,1MR,1Ros}.

\subsubsection{The function $x^p$}
First we find the fractional integral of the function $x^p$ using the
Riemann-Liouville definition, i.e.,  we have, with $a=0$,
\begin{eqnarray}
{{d^qx^p}\over{dx^q}}&=&{1\over\Gamma(-q)}{\int_0^x{{y^p}\over{(x-y)^{q+1}}}}dy
\;\;\;{\rm for}\;\;\;q<0.
\end{eqnarray}
With a change of variable, we get
\begin{eqnarray}
{{d^qx^p}\over{dx^q}}&=& {x^{p-q}\over\Gamma(-q)} 
\int_0^1 u^p (1-u)^{-q-1} du, \;\;\;{\rm for}\;\;\;q<0 \\
&=& {x^{p-q}\over\Gamma(-q)} B(p+1, -q),\;\;\;q<0,\; p>-1,
\end{eqnarray}
where $B(p,q)$ is the beta function. Therefore fractional integral of a $x^p$
is, for $p>-1$,
\begin{eqnarray}
{{d^qx^p}\over{dx^q}} = {\Gamma(p+1)\over\Gamma(p-q+1)} x^{p-q}, \;\;\;
q<0,\; p>-1. 
\end{eqnarray}
To find fractional derivative we use equation~(\ref{defRLd}) and obtain
\begin{eqnarray}
{{d^qx^p}\over{dx^q}}&=& {{d^n}\over{dx^n}}{{d^{q-n}x^p}\over{dx^{q-n}}}
\nonumber \\
&=& {{d^n}\over{dx^n}} \left[ {x^{p-q+n}\over\Gamma(n-q)} 
\int_0^1 u^p (1-u)^{n-q-1} du \right], \;\;\;{\rm for}\;\;\;0\leq q<n \nonumber\\
&=& {\Gamma(p+1)\over\Gamma(p-q+1)} x^{p-q}, \;\;\;
q\geq 0,\; p>-1. \label{xp}
\end{eqnarray}

The following limiting behavior of the abovestated formula may be noted, viz.,
\begin{eqnarray}
\lim_{x\rightarrow 0^+}{{d^qx^p}\over{dx^q}} = \left\{ \begin{array}{ll}
0 & \mbox{if $q<p$ or $q=p+n$} \\
{\Gamma(p+1)}  & \mbox{if $q=p$} \\
\infty & \mbox{otherwise}. \end{array} \right.
\end{eqnarray}
This observation will be of crucial importance later in the thesis.

\subsubsection{The exponential function $\exp(x)$}
We use the power series expansion
\begin{eqnarray}
\exp(x) =  \sum_{j=0}^\infty {x^j\over \Gamma(j+1)},
\end{eqnarray}
which is valid for all $x$, and differintegrate it term by term. Therefore
we get
\begin{eqnarray}
{{d^q\exp(x)}\over{dx^q}} &=& x^{-q} \sum_{j=0}^\infty {x^j\over \Gamma(j-q+1)}
\nonumber \\
&=& {\gamma^*(-q,x) \over x^q}
\end{eqnarray}
where $\gamma^*(a,b)$ is an incomplete gamma function~\cite{1AS}.

\subsubsection{The logarithmic function $\ln(x)$}
We use the Riemann-Liouville integral and get
\begin{eqnarray}
{{d^q\ln(x)}\over{dx^q}}&=&{1\over\Gamma(-q)}{\int_0^x{{\ln(y)}\over{(x-y)^{q+1}}}}dy,
\;\;\;{\rm for}\;\;\;q<0, \nonumber \\
&=& {x^{-q}\ln(x) \over \gamma(-q)} \int_0^1 {dv \over v^{q+1}} +
{x^{-q} \over \gamma(-q)} \int_0^1 {\ln(1-v)dv \over v^{q+1}}
\end{eqnarray}
where we have made use of the substitution $v=(x-y)/x$. Evaluating the above
integrals we get
\begin{eqnarray}
{{d^q\ln(x)}\over{dx^q}} = {x^{-q} \over \gamma(1-q)} [\ln(x) - \gamma 
- \psi(1-q)]
\end{eqnarray}
where $\psi(x)$ is defined by
\begin{eqnarray}
\psi(x) = {1\over \Gamma(x)} {d\Gamma(x) \over dx}.
\end{eqnarray}
Using equation~(\ref{defRLd}) this result can be extended to $q>0$.

\subsubsection{Trigonometric functions}
The differintegration of sine and cosine functions results in complex
hypergeometric functions and is given in~\cite{1OS}. The purpose of this section
is to note some relations, among fractional derivatives of sine and
cosine functions.

If we choose $f(x)=\sin x$ and $a=0$ in the equation~(\ref{defRLd2}) 
we get
\begin{eqnarray}
{{d^q\sin(x)}\over {d x^q}}={{d^{q-1}\cos(x)}\over{d x^{q-1}}}, 
\end{eqnarray}
where $0<q<1$. Similarly, when we choose $f(x)=\cos x$ and $a=0$ we get,
for $0<q<1$,
\begin{eqnarray}
{{d^q\cos(x)}\over {d x^q}}={x^{-q}\over \Gamma(-q+1)} -
{{d^{q-1}\sin(x)}\over{d x^{q-1}}}.
\end{eqnarray}
These relations will be of use later in our discussion.

\section{Fractals}

\subsection{Introduction}
Loosely speaking, fractals are sets or objects having non-integer
dimensions. Enormous body of literature already exists on this very active 
field~\cite{1Man,1Fal,1Edg,1Vic,1Fed}.
Conventionally, the dimension
of a object is usually a non-negative {\it integer}
(though there are situations where negative dimensions are also 
sensible~\cite{1Man2}), 
and it is defined through
 the number of coordinates needed for complete
specification of the given object.
Therefore, when one wonders about the possibility of existence of entities
which have {\it non-integer} dimension, one has to look for a different 
definition
of dimension which does not depend on the coordinates.

The intuition for such a definition comes from the way one measures length, 
area or volume
in practice. To make the basic idea of the definitions of fractional
dimensions clear we present this procedure explicitly.
In order to find the length of a smooth curve, one takes a
measuring stick of length say $\delta$ and makes tick marks at regular
intervals on the curve with the help of this stick. 
 Then the length $L$ of the curve is the product of $N(\delta)$,
the number of such marks, and $\delta$.
In order to make an finer estimate of the length, one makes $\delta$ smaller. 
One expects that as $\delta$ approaches a limit $0$ the quantity $L$ remains
bounded and approaches a limit. In similar manner in order to find area (volume) of a set it is covered with squares (cubes) of sides $\delta$ and the
number $N(\delta)$ of such squares (cubes) needed to cover a given set
is counted.
Then the area (volume) is defined as $A=\lim_{\delta \rightarrow 0}\delta^2 
N(\delta)$ 
($V=\lim_{\delta \rightarrow 0}\delta^3 N(\delta)$). Notice that the power of 
the $\delta$ above
corresponds to the dimension of the sets. 
Notice also that for a planar piece, the length is infinite,
the area is finite and the volume is
zero. The value of the dimension where the
content measure of the set jumps from infinity to zero is the characteristic
dimension of the set.
This intuition will provide the foundation
in the following formal definitions of fractional dimension.

\subsection{Definitions}
There exist numerous definitions of fractional dimension, each having its own advantages and disadvantages. We shall consider two important definitions
of dimension below. We shall follow the treatment of ref.~\cite{1Fal}.
Before going to the definitions let us list some desirable
properties, as discussed in~\cite{1Fal}, 
which a definition of dimension should have.
Let $E, F \subset I\!\!R^n$.
\begin{enumerate}
\item Monotonicity: If $E \subset F$ then $\mbox{dim}E \leq \mbox{dim}F$.

This is obvious since if we add some points to a given set its dimension
should not decrease.

\item Stability: $\mbox{dim}(E\bigcup F) = \max(\mbox{dim}E,\mbox{dim}F)$
\item Countable stability: $\mbox{dim}(\bigcup_{i=1}^{\infty}F_i) = 
\sup_{i\leq i < \infty} \mbox{dim}F_i$

The above two properties imply that if we combine a set with
other sets having lower
dimensions, the dimension of the resultant set does not change.

\item Geometric invariance: If $I\!\!R^n$ is transformed with a transformation 
such as a translation, rotation, similarity or affinity, the dimension of the
set $F\subset I\!\!R^n$ should not change.
\item Lipschitz invariance: If $f:F\rightarrow I\!\!R^m$ is a bi-lipschitz 
transformation, i.e.
\begin{eqnarray}
c_1|x-y| \leq |f(x)-f(y)| \leq c_2 |x-y|,\;\;\;\;x,y\in F, \nonumber
\end{eqnarray}
where $0<c_1\leq c_2 < \infty$, then $\mbox{dim}f(F) = \mbox{dim}F$.
\item Countable sets: If $F$ is finite or countable then $\mbox{dim}F=0$.

This is again expected since the dimension of the singleton set is zero and
from the countable stability any countable set should have dimension zero.

\item Open sets: If $F$ is an open subset of $I\!\!R^n$ then $\mbox{dim}F=n$.
\item Smooth manifolds: If $F$ is a smooth $m$-dimensional
manifold then $\mbox{dim}F=m$.
\end{enumerate}
Of course, not every definition of the dimension satisfies all the above 
properties.

\subsubsection{Hausdorff dimension}
Mathematically this is one of the most important definition of the dimension, 
since it is defined for any set. A major disadvantage of this definition is 
that it is not algorithmic.

Let us first define some quantities which are needed in the definition of
the Hausdorff dimension.
If $U$ is a non-empty subset of $I\!\!R^n$, the {\it diameter} of $U$ is defined as $|U| = \sup \{ |x-y| : x,y \in U \}$.
If $\{U_i\}$ is a countable collection
of sets of diameter at most $\delta$ that cover the given set $F$, i.e.
$F \subset \bigcup_{i=1}^{\infty} U_i$, we say that $\{U_i\}$ is a
$\delta$-{\it cover} of $F$.

Now we define a quantity, for any $\delta > 0$,
\begin{eqnarray}
{\cal{H}}^s_{\delta}(F) = \inf \left\{ \sum_{i=1}^\infty |U_i|^s :
\{U_i\} \mbox{is a $\delta$-cover of $F$} \right\}, \label{eq:Haus1}
\end{eqnarray}
where  $s$ is a non-negative 
number. In the above equation we consider all covers of $F$ by sets of
diameter at most $\delta$ and take the infimum of the sum of the $s$th powers
of the diameters. As $\delta$ decreases, the number of permissible covers
of $F$ in (\ref{eq:Haus1}) is reduced. Therefore, ${\cal{H}}^s_{\delta}(F)$
is a monotonic function, and approaches a 
limit (finite or $+\infty$) as $\delta \rightarrow 0$.
Now we define the Hausdorff measure by
\begin{eqnarray}
{\cal{H}}^s(F) = \lim_{\delta \rightarrow 0} {\cal{H}}^s_{\delta}(F).
\end{eqnarray}
This limit exists for any subset $F$ of $I\!\!R^n$. 

It can be 
proved~\cite{1Fal} that if ${\cal{H}}^s(F)$ is finite for $s=s_1$ then
${\cal{H}}^s(F)=0$ for $s>s_1$. This implies that
there exists a critical value of $s$ at which ${\cal{H}}^s(F)$ jumps from
$\infty$ to 0. This critical value is the Hausdorff dimension of $F$.

It is a customary~\cite{1Man} to normalize the Hausdorff measure by multiplying
it with
\begin{eqnarray}
\gamma(s) = {[\Gamma(1/2)]^s \over \Gamma(1+s/2)}, \label{eq:norm}
\end{eqnarray}
which is a $s$-dimensional volume of a ball of unit radius.

\subsubsection{Box dimension}
This is the most popular definition of dimension since it is easy to use
both mathematically and numerically. 

In order to define the box dimension consider the collection of cubes in the
$\delta$-coordinate mesh and $N_{\delta}(F)$ be the smallest number of such
cubes required to cover a non-empty bounded set. Then the lower and upper box dimensions of $F$ are respectively defined as
\begin{eqnarray}
\underline{\rm dim}_{\rm B}F = \underline{\lim}_{\delta \rightarrow 0}
{{\log  N_{\delta}(F)} \over{-\log{\delta}}}
\end{eqnarray}
\begin{eqnarray}
\overline{\rm dim}_{\rm B}F = \overline{\lim}_{\delta \rightarrow 0}
{{\log  N_{\delta}(F)} \over{-\log{\delta}}}.
\end{eqnarray}
The box dimension of $F$ is defined as
\begin{eqnarray}
{\rm dim_B}F=\underline{\rm dim}_{\rm B}F=\overline{\rm dim}_{\rm B}F
\end{eqnarray}
whenever the two limits are equal.

This definition of dimension satisfies most of the aforementioned desirable 
properties
of dimension, except that it is not countably stable and hence
countable sets may not have box dimension 0. For example, consider 
a set of rational numbers in the interval $[0,1]$. If one covers this set
by a partition with interval $\delta$ then the $N(\delta)=1/\delta$ and
hence the box dimension of the set is 1. But the set of rational numbers
is a countable set, i.e., it is a countable union of sets of dimension 
zero. Hence one would expect the set of rationals to have dimension 0.
 In spite of this paradoxical result this definition is widely
used by the physicists because of its simplicity, under the assumption that 
such cases are pathological in physical world.

It may be noted that in the definition of the box dimension we had essentially
covered the set with boxes of fixed size whereas in the Hausdorff
dimension we allow all sizes smaller than some  
$\delta$. This is the crucial difference between the two
definitions. We now see that because of this difference the Hausdorff
dimension of  the set of rationals in $[0,1]$ is 0. Let us cover the set of 
rationals in $[0,1]$ in following manner. Since this is a countable set we 
can label each rational number by a positive integer say $k$. Now cover
$k$th rational number by an interval of length $\delta / 2^k$.
Then the sum $\sum \delta^s/2^{ks}$ is bounded by $K\delta^s$ (for
some positive constant $K$) which goes to 0, for $s>0$, in the limit 
$\delta \rightarrow 0$. Hence the quantity ${\cal{H}}^s(F)$ is 0 for
any $s>0$.

\subsection{Examples}
We consider here two simple and well-known examples of fractal sets, i.e., 
the Cantor set and the von Koch curve. 
Graph of the Weierstrass function discussed in chapter~\ref{ch:3} provides
still another example.
Many more examples can similarly
be constructed. They are found, for example, in~\cite{1Fal,1Man,1Edg}.

\subsubsection{The middle-third Cantor set}
\begin{figure}[t] 
\begin{picture}(350,350)(-10,-300)
\put(0,0){\line(1,0){333}}\put(340,0){$E_0$}

\put(0,-50){\line(1,0){111}}
\put(222,-50){\line(1,0){111}}\put(340,-50){$E_1$}

\put(0,-100){\line(1,0){37}}
\put(74,-100){\line(1,0){37}}
\put(222,-100){\line(1,0){37}}
\put(296,-100){\line(1,0){37}}\put(340,-100){$E_2$}

\put(0,-150){\line(1,0){12.33}}
\put(24.66,-150){\line(1,0){12.33}}
\put(74,-150){\line(1,0){12.33}}
\put(98.66,-150){\line(1,0){12.33}}
\put(222,-150){\line(1,0){12.33}}
\put(246.66,-150){\line(1,0){12.33}}
\put(296,-150){\line(1,0){12.33}}
\put(320.66,-150){\line(1,0){12.33}}

\put(0,-200){\line(1,0){4.11}}
\put(8.22,-200){\line(1,0){4.11}}
\put(24.66,-200){\line(1,0){4.11}}
\put(32.88,-200){\line(1,0){4.11}}
\put(74,-200){\line(1,0){4.11}}
\put(82.22,-200){\line(1,0){4.11}}
\put(98.66,-200){\line(1,0){4.11}}
\put(106.88,-200){\line(1,0){4.11}}
\put(222,-200){\line(1,0){4.11}}
\put(230.22,-200){\line(1,0){4.11}}
\put(246.66,-200){\line(1,0){4.11}}
\put(254.88,-200){\line(1,0){4.11}}
\put(296,-200){\line(1,0){4.11}}
\put(304.22,-200){\line(1,0){4.11}}
\put(320.66,-200){\line(1,0){4.11}}
\put(328.88,-200){\line(1,0){4.11}}

\put(0,-250){\line(1,0){1.37}}
\put(8.22,-250){\line(1,0){1.37}}
\put(24.66,-250){\line(1,0){1.37}}
\put(32.88,-250){\line(1,0){1.37}}
\put(74,-250){\line(1,0){1.37}}
\put(82.22,-250){\line(1,0){1.37}}
\put(98.66,-250){\line(1,0){1.37}}
\put(106.88,-250){\line(1,0){1.37}}
\put(222,-250){\line(1,0){1.37}}
\put(230.22,-250){\line(1,0){1.37}}
\put(246.66,-250){\line(1,0){1.37}}
\put(254.88,-250){\line(1,0){1.37}}
\put(296,-250){\line(1,0){1.37}}
\put(304.22,-250){\line(1,0){1.37}}
\put(320.66,-250){\line(1,0){1.37}}
\put(328.88,-250){\line(1,0){1.37}}

\put(2.74,-250){\line(1,0){1.37}}
\put(10.96,-250){\line(1,0){1.37}}
\put(27.40,-250){\line(1,0){1.37}}
\put(35.62,-250){\line(1,0){1.37}}
\put(76.74,-250){\line(1,0){1.37}}
\put(84.96,-250){\line(1,0){1.37}}
\put(101.4,-250){\line(1,0){1.37}}
\put(109.62,-250){\line(1,0){1.37}}
\put(224.74,-250){\line(1,0){1.37}}
\put(232.96,-250){\line(1,0){1.37}}
\put(249.4,-250){\line(1,0){1.37}}
\put(257.62,-250){\line(1,0){1.37}}
\put(298.74,-250){\line(1,0){1.37}}
\put(306.96,-250){\line(1,0){1.37}}
\put(323.4,-250){\line(1,0){1.37}}
\put(331.62,-250){\line(1,0){1.37}}\put(340,-250){$F$}
\end{picture}
\caption{The middle third Cantor set}\label{fg:cantor}
\end{figure}
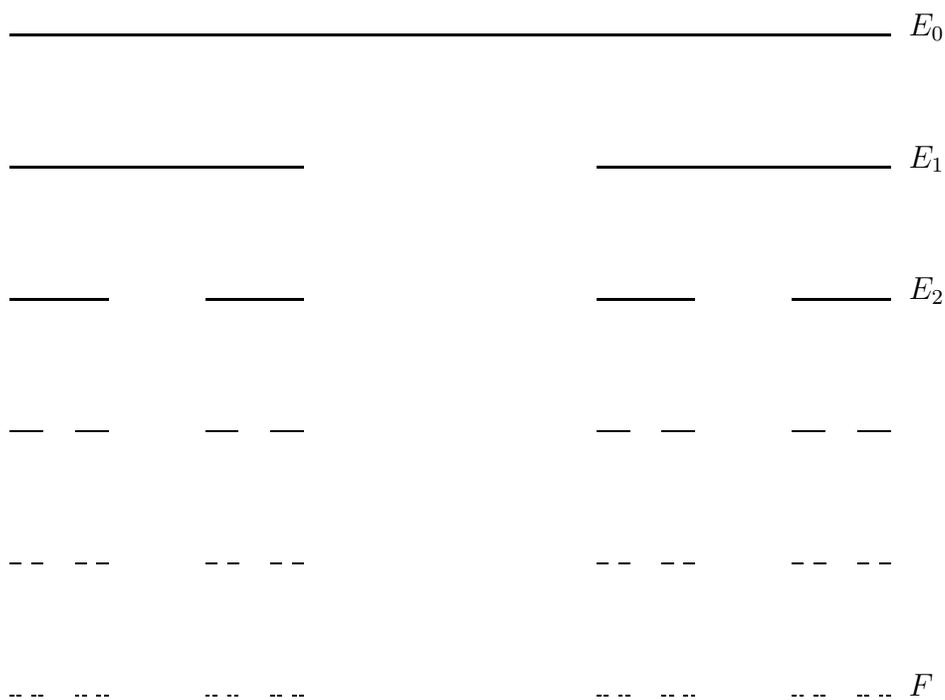

The middle-third Cantor set is constructed from a unit interval $[0,1]$
by a sequence of deletion operations. Let $E_0$ be the interval $[0,1]$.
Let $E_1$ be the set obtained by deleting the middle-third of $E_0$
(see figure~\ref{fg:cantor}). Therefore
the set $E_1$ consists of two intervals $[0,1/3]$ and $[2/3,1]$. The
set $E_2$ consisting of four intervals $[0,1/9]$, $[2/9,1/3]$, $[2/7,7/9]$,
$[8/9,1]$ is obtained by deleting middle thirds of the above two intervals.
We can continue in this way, deleting the middle thirds of each interval
in $E_{k-1}$ to obtain $E_k$, ad infinitum. The limiting set we get is called
the middle third cantor set. Notice that the set $E_k$ consists of $2^k$ 
intervals of length $3^{-k}$. That means we need $2^k$ intervals of length
$3^{-k}$ to cover the set at $k$th level. Therefore the box dimension
of the set is clearly $\log{2}/\log{3}$.

\subsubsection{The von Koch curve}
\begin{figure}
\begin{picture}(350,350)(-10,-300)
\put(0,0){\line(1,0){333}}\put(340,0){$E_0$}

\put(0,-100){\line(1,0){111}}
\put(111,-100){\line(3,5){55.5}}
\put(222,-100){\line(-3,5){55.5}}
\put(222,-100){\line(1,0){111}}\put(340,-100){$E_1$}

\put(0,-200){\line(1,0){37}}
\put(37,-200){\line(3,5){18.5}}
\put(74,-200){\line(-3,5){18.5}}
\put(74,-200){\line(1,0){37}}
\put(111,-200){\line(3,5){18.5}}
\put(129.5,-169.17){\line(-3,5){18.5}}
\put(148,-138.33){\line(-1,0){37}}
\put(166.5,-107.5){\line(-3,-5){18.5}}
\put(166.5,-107.5){\line(3,-5){18.5}}
\put(185,-138.33){\line(1,0){37}}
\put(203.5,-169.17){\line(3,5){18.5}}
\put(222,-200){\line(-3,5){18.5}}
\put(222,-200){\line(1,0){37}}
\put(259,-200){\line(3,5){18.5}}
\put(296,-200){\line(-3,5){18.5}}
\put(296,-200){\line(1,0){37}}\put(340,-200){$E_2$}

\end{picture}
\caption{The construction of von Koch curve}\label{fg:vonkoch}
\end{figure}
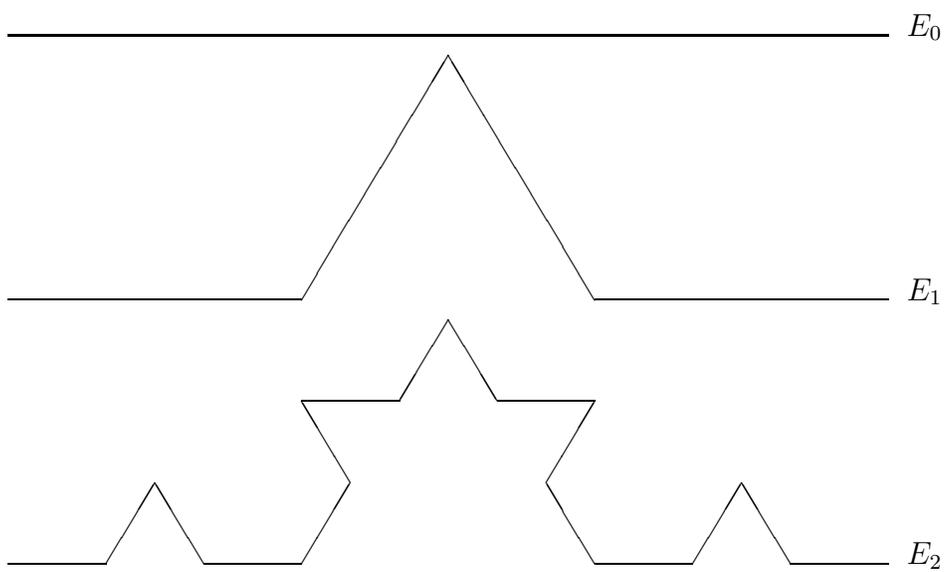

In order to construct the von Koch curve we again start with the unit interval
$[0,1]$. Now we remove the middle third segment of this interval and
instead of throwing it off we replace it, as shown in figure~\ref{fg:vonkoch}, 
by the other two sides of the
equilateral triangle based on the removed segment. In this way we get the
set $E_1$ which consists of four segments. By applying the same procedure 
to each segment of $E_1$ we obtain the set $E_2$, and so on. The limiting
set we get is called the von Koch curve. It may be noticed that 
the set $E_k$ consists of $4^k$ segments of length $3^{-k}$. Therefore
the dimension of the von Koch curve is $\log{4}/\log{3}$.

These two examples of fractals, though simple, display many typical
characteristics of fractals. The set $F$ in both these examples is
self-similar, i.e., it contains copies of itself on all length scales.
The set $F$ contains intricate detailed structure at arbitrarily small
scales. In spite of this, the set is easy to construct using a recursive
procedure. More importantly, the geometry of $F$ is not easily described
in standard notion, i.e., it is not the locus of points that satisfy
some simple geometric condition, or it is not a solution of any simple equation.

\subsection{Occurrences of Fractals}
The notion of fractals has found innumerable applications 
in pure as well as in applied
sciences. The present thesis itself contains lot of such applications. 
We have already mentioned some examples of fractals in the beginning of
the thesis. In the following we discuss some additional physical applications.
Many more applications may be found in
references~\cite{1Vic,1Fed,1BH,1BS}.

\subsubsection{Kinetic Aggregation}
The diffusion-limited aggregation (DLA) model~\cite{1BS,1BH2} 
is a simple model of a fractal,
generated by diffusion of particles in the following manner. 
One starts with a `seed'
particle which is fixed at a point. A second particle starts on the border of a
circle around this point and performs a random walk. When it comes near
the seed it sticks to it and a cluster of two particles is formed. Then a third
particle performs a similar random walk and forms a cluster of three particles.
 This procedure
is repeated many times, thereby giving rise to a large cluster having the shape
of a fractal. 
DLA models many 
phenomena in nature, such as, viscous fingering, dielectric breakdown,
snowflake growth, etc.

\subsubsection{Biological systems}
There is abundance of fractal structures and processes in biological systems.
Human lungs, branching of trees, root system in plants, etc, have a selfsimilar
branching structure which is typical of fractals~\cite{1Bul}. 
The abovementioned 
DLA model has also been
used to understand the shape of a neuron, growth of bacterial colonies, etc.
Long range correlations, which are believed to give rise to fractal geometry,
are found in DNA sequences, human heart beats etc.

\subsubsection{Percolation}
Consider a large square lattice, where each site is occupied randomly with 
probability $p$ and empty with probability $1-p$. At low values of $p$,
the occupied sites  form small clusters. As one increases $p$ there
exists a threshold $p_c$ at which a large (macroscopic) cluster appears and
which connects opposite edges of the lattice. This is called the
percolation cluster. This percolation cluster is selfsimilar in nature
and hence has fractal structure.

\section{Applications to the study of scaling in physical systems}
Recent work~\cite{1Hil2,1GR,1RG,1GN,1Zas}
shows that the fractional calculus formalism is useful
in dealing with scaling phenomena. The purpose of this section is to present
a brief review of  a few developments in this area.

\subsection*{Fractional Brownian Motion}
 Mandelbrot and Van Ness~\cite{1MV} have 
used fractional integrals to formulate fractal processes such as fractional
Brownian motion.
The fractional Brownian motion is described by the probability distribution
 $B_H(t)$ defined by, 
\begin{eqnarray}
B_H(t)-B_H(0) = {1\over\Gamma(H+1)} \int_{-\infty}^t K(t-t') dB(t'),\;\;\;
0<H<1,
\end{eqnarray}
where $B(t)$ is an ordinary Gaussian random process with zero mean and
unit variance, and with $K(t-t')$ such that
\begin{eqnarray}
K(t-t') = \left\{ 
\begin{array}{ll}
(t-t')^{H-1/2} & 0\leq t' <t \\
\{(t-t')^{H-1/2}-(-t)^{H-1/2}\} & t' < 0.
\end{array}
\right.
\end{eqnarray}
The use of the fractional integral kernel is apparent.
 Recently  Sebastian~\cite{1Seb} has given a path integral
representation for fractional Brownian motion whose measure contains fractional
derivatives of paths.

\subsection*{Fractional equations for a class of L\'evy type probability
densities}
L\'evy flights~\cite{1BG} and L\'evy walks~\cite{1SWK,1SZK} 
have found applications in various branches
of physics \cite{1Shl2} ranging from 
fluid dynamics~\cite{1SWS} to polymers~\cite{1OBLU}. 
In~\cite{1Non} T. F. Nonnenmacher considered a class of normalized one-sided 
L\'evy type probability densities, which provides a distribution for the
 lengths of  jumps of the random walker, given by
\begin{eqnarray}
f(x)={a^{\mu}\over{\Gamma(\mu)}} x^{-\mu-1} \exp(-a/x) \;\;\;a>0,\;x>0.
\end{eqnarray}
It is clear that for large $x$,
\begin{eqnarray}
f(x) \sim x^{-\mu-1}  \;\;\;\mu>0,\;x>0,
\end{eqnarray}
where $\mu$ is the L\'evy index. In~\cite{1Non} it was
 shown that these L\'evy-type
probability densities satisfy a fractional integral equation of the form
\begin{eqnarray}
x^{2q}f(x) = a^q {d^{-q}f(x)\over{dx^{-q}}},
\end{eqnarray}
or equivalently, a fractional differential equation given by
\begin{eqnarray}
{d^{n-\mu}f(x)\over{dx^{n-\mu}}} = a^{-\mu}{d^n\over{dx^n}}(x^{2\mu}f(x)).
\end{eqnarray}
Here, $n=1$ if $0<\mu<1$ and $n=2$ if $1<\mu<2$, etc. 
An interesting observation is that the L\'evy index is related
to the order of the fractional integral or differential operator.

\subsection*{Fractional equations for physical processes}
Modifications of standard equations governing physical processes such as
 diffusion equation, wave equation and Fokker-Planck equation have been
suggested~\cite{1GR,1RG,1Wys,1SW,1Jum}, which incorporate fractional derivatives with respect to time.
In refs.~\cite{1GR,1RG} a fractional diffusion equation has been
proposed for the diffusion on fractals. Asymptotic solution of this
equation coincides with the result obtained numerically.
Fogedby~\cite{1Fog} has considered a generalization of the
Fokker-Planck equation involving addition of 
a fractional gradient operator (defined as the Fourier
transform of $-k^{\mu}$) to the usual Fokker-Planck equation
 and performed a renormalization group analysis.

\subsection*{Transport in chaotic Hamiltonian systems}
Recently Zaslavsky~\cite{1Zas} showed that chaotic Hamiltonian  dynamics,
under certain circumstances, can be described by a fractional generalization of 
the Fokker-Planck-Kolmogorov equation, which is defined by two fractional
critical exponents $(\alpha , \beta)$ respectively for the space and
time derivatives of the distribution function.
With certain assumptions, the following equation  was derived for the
transition probabilities $P(x,t)$:
\begin{eqnarray}
{\partial^{\beta}P(x,t)\over{\partial t^{\beta}}} = {\partial^{\alpha}\over
{\partial(-x)^{\alpha}}}(A(x)P(x,t)) + {1\over 2} {\partial^{2\alpha}\over
{\partial(-x)^{2\alpha}}}(B(x)P(x,t))
\end{eqnarray}
where 
\begin{eqnarray}
A(x)=\lim_{\Delta t \rightarrow 0} {A'(x;\Delta t)\over{(\Delta t)^{\beta}}}
\end{eqnarray}
and
\begin{eqnarray}
B(x)=\lim_{\Delta t \rightarrow 0} {B'(x;\Delta t)\over{(\Delta t)^{\beta}}}
\end{eqnarray}
with $A'$ and $B'$ being $\alpha^{th}$ and $2\alpha^{th}$ moments respectively.
The exponents $\alpha$ and $\beta$ have been related to anomalous transport
exponent.

\subsection*{Some more instances}

\begin{itemize}
\item Hilfer~\cite{1Hil1,1Hil2,1Kol} has generalized the Eherenfest's 
classification
of phase transition by analytically continuing the order
which is equivalent to using the fractional derivative.  

\item Some recent papers~\cite{1Non,1GR,1RG,1PZ} indicate a connection 
between fractional calculus
and fractal structure~\cite{1Man,1Fal} or fractal 
processes~\cite{1MV,1GN,1Shl1}.
\item W. G. Gl\"ockle and T. F. Nonnenmacher~\cite{1GN} have formulated fractional
differential equations for some relaxation processes which are essentially
fractal time~\cite{1Shl1} processes.
\end{itemize}

We note that the most of the applications  reviewed in this section
deal with {\it asymptotic scaling only}. As noted in~\cite{1Gio}, in the context
of Schr\"odinger equation containing fractional operators, these fractional
differential equations are approximations as they do not 
completely take into account 
the {\it local scaling} property of the fractals. 
Also, in some of the examples above the dynamics is governed by 
 non-local  equations owing to the fact that the fractional time
derivatives used are non-local. This point has also been emphasized
in~\cite{1Hil3}. In the later part of the thesis, we develop a new notion
which takes care of these problems.

\section{Weyl fractional differentiability}
The purpose of this section is to review 
briefly, for the sake of completeness, 
the classic works of Stein, Zygmund and others 
\cite{1Zyg,1Ste,1SZ,1Wel},
who use Weyl fractional calculus in the analyses of irregular functions.

\subsection{H\"older exponent} \label{se:holder}
Let $f(x)$ be defined in a closed interval $I$ of real line, and let
\begin{eqnarray}
\omega(\delta)\equiv\omega(\delta : f) = \sup \vert f(x)-f(y) \vert
\;\;\;{\rm for }\;\;\;x,y \in I \;\;\;\vert x-y \vert \leq \delta
\end{eqnarray}
The function $\omega(\delta)$ is called the modulus of continuity of $f$.
If for some $\alpha > 0$ we have $\omega(\delta) \leq C {\delta}^{\alpha}$
with $C$ independent of $\delta$, then $f$ is said to satisfy H\"older 
(In old literature it is also called Lipschitz)
condition of order $\alpha$. These functions define a class 
${\Lambda}_{\alpha}$ of functions. Notice that with this definition of the 
modulus of continuity
only the case $0<\alpha \leq 1$ is interesting, since
if $\alpha>1$ for all $x$ of $I$, 
$f'(x)$ is zero everywhere and the function is constant.
This definition has been generalized (see~\cite{1JM}) 
to one where the case $\alpha>1$ is also nontrivial.
The idea is to subtract an appropriate polynomial from the function $f$. 
\begin{eqnarray}
\tilde{\omega}(\delta)\equiv\tilde{\omega}(\delta : f) = \sup \vert f(x)-P(x-y) \vert
\;\;\;{\rm for }\;\;\;x,y \in I \;\;\;\vert x-y \vert \leq \delta
\end{eqnarray}
where $P$ is the only polynomial of smallest degree which gives the 
smallest order of magnitude for $\tilde{\omega}$.
Recently, this definition was extensively used~\cite{1MBA,1Jaf} to characterize
 the velocity field of a turbulent fluid. 

\subsection{Summary of classical results}
Following Welland~\cite{1Wel} we introduce
\begin{defn}
$f$ is said to have an $\alpha$th derivative, where $0\leq k < \alpha < k+1$
for integer $k$,
 if $D^{-\beta}f$ (in the Weyl sense), 
with $\beta = k+1-\alpha$, has  $k+1$ Peano
derivatives at $x_0$. That is, there exists a polynomial $P_{x_0}(t)$ of 
degree $\leq k+1$ s. t.
\begin{eqnarray}
(D^{-\beta}f)(x_0+t)-P_{x_0}(t) = o(\vert t \vert^{k+1}) \;\;\;\;\;t\rightarrow 0.
\end{eqnarray}
Further if 
\begin{eqnarray}
\{{1\over \rho}\int_{-\rho}^{\rho} \vert (D^{-\beta}f)(x_0+t)-P_{x_0}(t) \vert
^p dt \}^{1/ p} = o(\rho^{k+1}) \;\;\;\;\; \rho \rightarrow 0 \;\;\;(1\leq p
< \infty)
\end{eqnarray}
$f$ is said to have an $\alpha$th derivative in the $L^p$ sense.
\end{defn}
It is clear that this definition of fractional differentiability is {\it not
local}. Particularly the behavior of function at $-\infty$ also plays a crucial
role. The main results can be stated using this notion of
differentiability and involve the classes $\Lambda^p_{\alpha}$ and
$N^p_{\alpha}$ which are given by the following definitions~\cite{1SZ,1Wel}.
\begin{defn}
If there exists a polynomial $Q_{x_0}(t)$ of degree $\leq k$ s. t.
$f(x_0+t)-Q_{x_0}(t) = O(\vert t \vert^{\alpha})$ as $t\rightarrow 0$
then $f$ is said to satisfy the condition $\Lambda_{\alpha}$, and if
\begin{eqnarray}
\{{1\over \rho}\int_{-\rho}^{\rho} \vert f(x_0+t)-Q_{x_0}(t) \vert
^p dt \}^{1/ p} = O(\rho^{\alpha}) ,\;\;\; \rho \rightarrow 0 \;\;\;(1\leq p
< \infty)
\end{eqnarray}
$f$ is said to satisfy the condition $\Lambda_{\alpha}^p$.
\end{defn}
\begin{defn}
$f$ is said to satisfy the condition $N_{\alpha}^p$ if for some $\rho > 0$
\begin{eqnarray}
{1\over \rho}\int_{-\rho}^{\rho} {{\vert f(x_0+t)-Q_{x_0}(t) \vert
^p}\over{\vert t \vert^{1+p\alpha}}} dt < \infty.
\end{eqnarray}
\end{defn}
We are now in a position to state the main results in the form of 
the following  four theorems.
First two of these~\cite{1SZ,1Wel} state the condition under which the 
fractional derivative of
a function exists.
\begin{thm}
Suppose that $f$ satisfies the condition $\Lambda_{\alpha}$ at every point 
of a set $E$ of positive measure. Then $D^{\alpha}f(x)$ exists almost everywhere
in $E$ if and only if $f$ satisfies condition  $N_{\alpha}^2$ almost everywhere
in $E$.
\end{thm}
\begin{thm}
The necessary and sufficient condition that $f$ satisfies the condition
$N_{\alpha}$ almost everywhere in a set $E$ is that $f$ satisfies the condition
$\Lambda^2_{\alpha}$ and $D^{\alpha}f$ exists in the $L^2$ sense, almost
everywhere in this set.
\end{thm}
The following results~\cite{1Zyg} tell us how the class of a
function changes when an operation of fractional differentiation is performed.
\begin{thm}
Let $0\leq \alpha < 1$, $\beta > 0$ and suppose that $f\in \Lambda_{\alpha}$.
Then  $D^{-\beta}f \in \Lambda_{\alpha+\beta}$ if $\alpha+\beta < 1$.
\end{thm}
\begin{thm}
Let $0<\gamma < \alpha < 1$,.
Then  $D^{\gamma}f \in \Lambda_{\alpha-\gamma}$ if $f\in \Lambda_{\alpha}$.
\end{thm}

Despite their merits, these results are not really suitable and adequate
to obtain information regarding irregular behavior of functions and 
H\"older exponents. We observe that the Weyl definition involves highly
nonlocal information and hence is somewhat unsuitable for treatment of
local scaling behavior. In the next section we introduce a more
appropriate definition.

\chapter{Local Fractional Derivatives} \label{ch:2}

\section{Motivation}
The definition of the fractional derivative was discussed in the last chapter.
These derivatives differ in some aspects from  integer order derivatives.
In order to see this, one may note, from equation~(\ref{defRLd}), 
that except when $q$ is a positive integer, the $q^{th}$ derivative
is nonlocal as it depends on the lower limit `$a$'.
The same feature is also shown by other definitions.
However, we wish to
study local scaling properties and hence we need to modify this definition
accordingly.
Secondly from equation~(\ref{xp}) it is clear that the fractional derivative of a constant function is not zero. Therefore adding a constant to a function 
alters the value of the fractional derivative. This is an undesirable
property of the fractional derivatives to study fractional differentiability.
While constructing {\it local fractional derivative} operator,
we have to correct for these two features. 
This forces one to choose the lower limit as well as the additive
constant before hand. The most natural choices are as follows.
1) We subtract, from the function, the value of the function at the point where
 we want to study the local scaling property. 
This makes the value of the function
zero at that point, canceling the effect of any constant term.
 2) Natural choice of a lower limit will again be
that point itself, where we intend to examine the local scaling.

In the following section we formally introduce the concept of local
fractional derivative.

\section{Definition}

\begin{defn} \label{def:lfd}
If, for a function $f:[0,1]\rightarrow I\!\!R$, the  limit 
\begin{eqnarray}
I\!\!D^qf(y) = 
{\lim_{x\rightarrow y} {{d^q(f(x)-f(y))}\over{d(x-y)^q}}},
\;\;\;\;0<q\leq 1\label{defloc}
\end{eqnarray}
exists and is finite, then we say that the {\it local fractional derivative} (LFD) 
of order $q$ (denoted by $I\!\!D^qf(y)$), at $y$, 
exists. 
\end{defn}
This defines the LFD for $0<q\leq 1$. It was first introduced
in~\cite{2KG1}, 
and later generalized~\cite{2KG2} to include all positive values of $q$ as
follows.
\begin{defn} 
If, for a function $f:[0,1]\rightarrow I\!\!R$, the  limit 
\begin{eqnarray}
I\!\!D^qf(y) =  {\lim_{x\rightarrow y}}
{{d^q(f(x)-\sum_{n=0}^N{f^{(n)}(y)\over\Gamma(n+1)}(x-y)^n)}
\over{[d(x-y)]^q}} \label{deflocg}
\end{eqnarray}
exists and is finite,
where $N$ is the largest integer for which $N^{th}$ derivative of $f(x)$ at
$y$ exists and is finite, then we say that the {\it local fractional
derivative} (LFD) of order $q$ $(N<q\leq N+1)$, at $x=y$, 
exists. 
\end{defn}
We subtract the Taylor series term in the above
definition for the same reason as one subtracts $f(y)$ in the 
definition~\ref{def:lfd}. We do this to supress any regular behavior
that may mask the local singularity. 

\begin{defn}
The {\it critical order} $\alpha$, at $y$, of a function $f$ is
\begin{eqnarray}
\alpha(y) = \sup \{q \vert I\!\!D^{q'}f(y), \;q'<q, \;\mbox{exists}\}. \nonumber
\end{eqnarray}
\end{defn}

Sometimes it is essential to distinguish between limits, and hence the critical
order, taken from above and below. In that case we define
\begin{eqnarray}
I\!\!D_{\pm}^qf(y) =  {\lim_{x\rightarrow y^{\pm}}}
{{d^q(f(x)-\sum_{n=0}^N{f^{(n)}(y)\over\Gamma(n+1)}(x-y)^n)}
\over{[d\pm(x-y)]^q}}. \label{deflocg+}
\end{eqnarray}
We will assume $I\!\!D^q = I\!\!D_{+}^q$ unless mentioned otherwise.

As an example consider the function $f(x)=|x|^\alpha$, $\alpha \geq 0$. 
The critical order of this function from above
at origin is $\alpha$ when $\alpha$ is noninteger and is
$\infty$ when $\alpha$ is an integer.

The local fractional derivative that we have defined above reduces to the usual
derivatives of integer order when $q$ is a positive integer. 
In order to see this point 
for $q=1$ we consider equation~(\ref{defloc}). 
Since the Riemann-Liouville fractional
derivative on RHS of equation~(\ref{defloc}) 
reduces to ordinary first derivative when $q=1$ (This follows from the relation
\begin{eqnarray}
{{d}\over{dx}}{{d^qf(x)}
\over{[d(x-y)]^q}}={{d^{q+1}f(x)}
\over{[d(x-y)]^{q+1}}} \nonumber
\end{eqnarray}
and setting $q=0$. For details, see the book by Oldham and Spanier~\cite{2OS}, 
page 50. See also ref.~\cite{2MR}), we get
\begin{eqnarray}
I\!\!D^1f(y) =  {\lim_{x\rightarrow y}} {{d[f(x)-f(y)]}\over{dx}}.
\end{eqnarray}
Now since $f(y)$ is constant its derivative is zero. Therefore after
taking the limit we get
\begin{eqnarray}
I\!\!D^1f(y) =  {df(y)\over{dy}}.
\end{eqnarray}
For $N<q\leq N+1$ the definition of local fractional derivative is
\begin{eqnarray}
I\!\!D^qf(y) =  {\lim_{x\rightarrow y}}
{{d^q[f(x)-\sum_{i=0}^N{f^{(i)}(y)\over\Gamma(i+1)}(x-y)^i]}
\over{[d(x-y)]^q}} 
\end{eqnarray}
Now, if $q=n$ (i.e. $N=n-1$ in view of the above range of $q$), we get
\begin{eqnarray}
I\!\!D^nf(y) =  {\lim_{x\rightarrow y}}
{{d^n[f(x)-\sum_{i=0}^N{f^{(i)}(y)\over\Gamma(i+1)}(x-y)^i]}
\over{dx^n}}.
\end{eqnarray}
Again since $y$ is a constant
\begin{eqnarray}
I\!\!D^nf(y) =  {{d^nf(y)}\over{dy^n}}.
\end{eqnarray}
Therefore, for $q=n$, local fractional derivative reduces to $n$th order
derivative.

From this it is clear that in our construction local fractional derivatives 
generalize
the usual derivatives to fractional order keeping the local nature
of the derivative operator intact, in contrast to other definitions
of fractional derivatives. The local nature of the operation of
derivation is crucial at many places, for instance, in studying
differentiable structure of complicated manifolds, studying evolution
of physical systems locally, etc. The virtue of such a local quantity
will be evident in the following section
where we show that the local fractional derivative 
 appears naturally in the fractional Taylor expansion.
This will imply that the LFDs are not introduced 
in an ad hoc manner merely to satisfy the
two conditions mentioned in the beginning, but
they have their own importance.

\section{Local fractional Taylor expansion}\label{se:lfte}
An interesting consequence of the above definitions is that the LFDs 
 appear naturally in the fractional Taylor expansion.

\subsubsection{Derivation of local fractional Taylor expansion}
We follow the usual procedure to derive Taylor expansion with a 
remainder~\cite{2CJ}.
In order to derive local fractional Taylor expansion, let
\begin{eqnarray}
F(y,x-y;q) = {d^q(f(x)-f(y))\over{[d(x-y)]^q}}.
\end{eqnarray}
It is clear that
\begin{eqnarray}
I\!\!D^qf(y)=F(y,0;q)
\end{eqnarray}
Now, for $0<q\leq 1$,
\begin{eqnarray}
f(x)-f(y)&=& {d^{-q}\over{[d(x-y)]^{-q}}}{d^{q}\over{[d(x-y)]^{q}}}f \nonumber\\
& =& {1\over\Gamma(q)} \int_0^{x-y} {F(y,t;q)\over{(x-y-t)^{-q+1}}}dt\\
&=& {1\over\Gamma(q)}[F(y,t;q) \int (x-y-t)^{q-1} dt]_0^{x-y} \nonumber\\
&&\;\;\;\;\;\;\;\;+ {1\over\Gamma(q)}\int_0^{x-y} {dF(y,t;q)\over{dt}}{(x-y-t)^q\over{q}}dt
\end{eqnarray}
provided the last term exists. Thus
\begin{eqnarray}
f(x)-f(y)&=& {I\!\!D^qf(y)\over \Gamma(q+1)} (x-y)^q \nonumber\\
&&\;\;\;\;\;\;\;\;+ {1\over\Gamma(q+1)}\int_0^{x-y} {dF(y,t;q)\over{dt}}{(x-y-t)^q}dt\label{taylor}
\end{eqnarray}
i.e.
\begin{eqnarray}
f(x) = f(y) + {I\!\!D^qf(y)\over \Gamma(q+1)} (x-y)^q + R_q(x,y) \label{taylor2}
\end{eqnarray}
where $R_q(x,y)$ is a remainder given by
\begin{eqnarray}
R_q(x,y) = {1\over\Gamma(q+1)}\int_0^{x-y} {dF(y,t;q)\over{dt}}{(x-y-t)^q}dt
\end{eqnarray}
Equation (\ref{taylor2}) is a fractional Taylor expansion of $f(x)$ involving
only the lowest and the second leading terms. Using the general definition
of LFD and following similar steps one arrives at the fractional Taylor
expansion for $N<q\leq N+1$ (provided $I\!\!D^q$ exists), given by,
\begin{eqnarray}
f(x) = \sum_{n=0}^{N}{f^{(n)}(y)\over{\Gamma(n+1)}}(x-y)^n
 + {I\!\!D^qf(y)\over \Gamma(q+1)} (x-y)^q + R_q(x,y) \label{taylorg}
\end{eqnarray}
where 
\begin{eqnarray}
R_q(x,y) = {1\over\Gamma(q+1)}\int_0^{x-y} {dF(y,t;q,N)\over{dt}}{(x-y-t)^q}dt
\end{eqnarray}

We note that the local fractional derivative (not just fractional derivative)
as defined above
 provides
the coefficient $A$ in the approximation
of $f(x)$ by the function $f(y) + A(x-y)^q/\Gamma(q+1)$, for $0<q<1$, 
in the vicinity of $y$. 
 We further note that the terms
on the RHS of eqn(\ref{taylor}) are nontrivial and finite only in the case $q=\alpha$.
Osler in ref.\cite{2Osl} has constructed fractional Taylor
series using usual (not local in the sense above) fractional derivatives. 
His results are, however, applicable to analytic functions and cannot be 
used for non-differentiable scaling functions directly. Furthermore, Osler's
formulation involves terms with negative $q$ also and hence is not suitable
for approximating schemes.

Let us consider the function $f(x)= x^\alpha$, $x,\alpha\geq 0$. Then 
$I\!\!D^\alpha f(0) = \Gamma(\alpha +1)$ and using equation~(\ref{taylorg})
at $y=0$ we get $f(x) = x^\alpha $ since the remainder term turns out to
be zero.

\subsubsection{Geometrical interpretation of LFD}
It is well known~\cite{2MR} that 
one can not attach any geometrical interpretation to
the conventional fractional derivatives as one does for ordinary derivatives.
 Whereas, the local fractional Taylor expansion of section~\ref{se:lfte}
suggests  a possibility of such an interpretation for LFDs.
In order to see this note that when $q$ is set equal to
unity in the equation~(\ref{taylor2}) one gets
the equation of the tangent. 
It may be recalled that all the curves passing through a point $y$ and having same tangent
form an equivalence class (which is modeled by a linear behavior). 
Analogously all the functions (curves) with the same critical order $\alpha$
and the same $I\!\!D^{\alpha}$
will form an equivalence class modeled by $x^{\alpha}$ 
(If $f$ differs from $x^{\alpha}$ by a logarithmic correction then 
terms on RHS of eqn(\ref{taylor})
do not make sense like in the ordinary calculus). 
This is how one may
generalize the geometric interpretation of derivatives in terms of `tangents'.  
This observation is useful when one wants to approximate an irregular 
function by a piecewise smooth (scaling) function.

\section{Generalization to higher dimensional functions} \label{higherd}
The definition of the Local fractional derivative can be generalized~\cite{2KG3}
for higher dimensional functions in the following manner.

Consider a function $f: I\!\!R^n \rightarrow I\!\!R$. We define
\begin{eqnarray}
\Phi({\bf y},t) = f({\bf y}+{\bf v}t) - f({\bf y}),\;\;\;
{\bf v} \in I\!\!R^n,\;\;\;t\in I\!\!R.
\end{eqnarray}
Then the directional-LFD of $f$ at ${\bf y}$ 
of order $q$, $0<q<1$, in the direction ${\bf v}$ is given 
(provided it exists) by
\begin{eqnarray}
I\!\!D^q_{\bf v}f({\bf y}) = {d^q{\Phi}({\bf y},t) \over{dt^q}}\vert_{t=0} 
\label{defloch}
\end{eqnarray}
where the RHS involves the usual fractional 
derivative of equation (\ref{defRLd}). The directional LFDs along the unit vector 
${\bf e}_i$ will be called $i^{\rm th}$ partial-LFD.

\section{Some remarks}
\begin{enumerate}

\item We would like to point out that 
there is a multiplicity of definitions of fractional derivatives.
The use of Riemann-Liouville
 definition, and other equivalent definitions such as Grunwald's  
definition are suitable for our purpose.
 The other definitions of fractional derivatives which
do not allow control over both the limits, such as Wyel's definition or definition
using Fourier transforms, are not suitable since
it would not be possible to retrieve local nature of
differentiability property which is essential for study of
local behavior. Also, the important difference between our work and
the work of \cite{2GR,2GN} is that while we are trying to study the local scaling behavior these works apply to asymptotic scaling properties.

\item It is interesting to note that the same definition of
LFD can be used for negative values of the critical order between -1 and 0.
For this range of critical orders $N=-1$ and the sum in equation 
(\ref{deflocg}) is empty. As a result the expression for LFD becomes
\begin{eqnarray}
I\!\!D^qf(y) = 
{\lim_{x\rightarrow y} {{d^qf(x)}\over{[d(x-y)]^q}}}
\end{eqnarray}
An equivalence between the critical order and the H\"older exponent,
for positive values of critical order, will be proved in chapter~\ref{ch:2}.
Here we would like to point out that the
negative H\"older exponents do arise in real physical situation of 
turbulent velocity field (see~\cite{2Eyi,2Jaf} and references therein).

\item Another way of generalizing the LFD to the values of critical order
beyond 1 would have been to write it as
\begin{eqnarray}
I\!\!D^qf(y) = 
{\lim_{x\rightarrow y} {{d^q(f^{(N)}(x)-f^{(N)}(y))}\over{[d(x-y)]^q}}}
\end{eqnarray}
But the existence of $N^{th}$ derivative of $f$ at $x$ may not be 
guaranteed in general. Such a situation may arise in the case of 
multifractal functions to be treated in chapter~\ref{ch:4}.

\end{enumerate}

\chapter{Fractional Differentiability of Nowhere Differentiable Functions}
\label{ch:3}

\section{Introduction to nowhere differentiable functions} \label{se:1ch3}
It is well known that a continuous function may not be differentiable at 
some point (e.g., $f(x) = |x|$ is not differentiable at $x=0$). 
But, as mentioned in Chapter 1, till the last century
it was generally believed that any continuous function must be differentiable
atleast at one point. The example of the Weierstrass continuous but nowhere 
differentiable function was a counterpoint to this belief. 
Weierstrass constructed the 
function
\begin{eqnarray}
W(t) = \sum_{k=1}^{\infty} a^k \cos{b^kt}, \label{eq:Wgen}
\end{eqnarray}
where $0<a<1<b$ and $ab>1$ (see \cite{3Har}). While he himself 
and others did independently prove that this
function is nowhere differentiable for some of these values of $a$ and $b$,
it was Hardy~\cite{3Har} who for the first time showed 
this for all such $a$ and $b$.

We shall consider the following form of Weierstrass function 
\begin{eqnarray}
W_{\lambda}(t) = \sum_{k=1}^{\infty} {\lambda}^{(s-2)k} 
\sin{\lambda}^kt,\;\;\;\;
\lambda>1\;\;\;1<s<2, \label{eq:Wsp}
\end{eqnarray}
where we have chosen $b=\lambda$ and $a={\lambda}^{(s-2)k}$.
Note that $W_{\lambda}(0)=0$. 

\begin{figure}
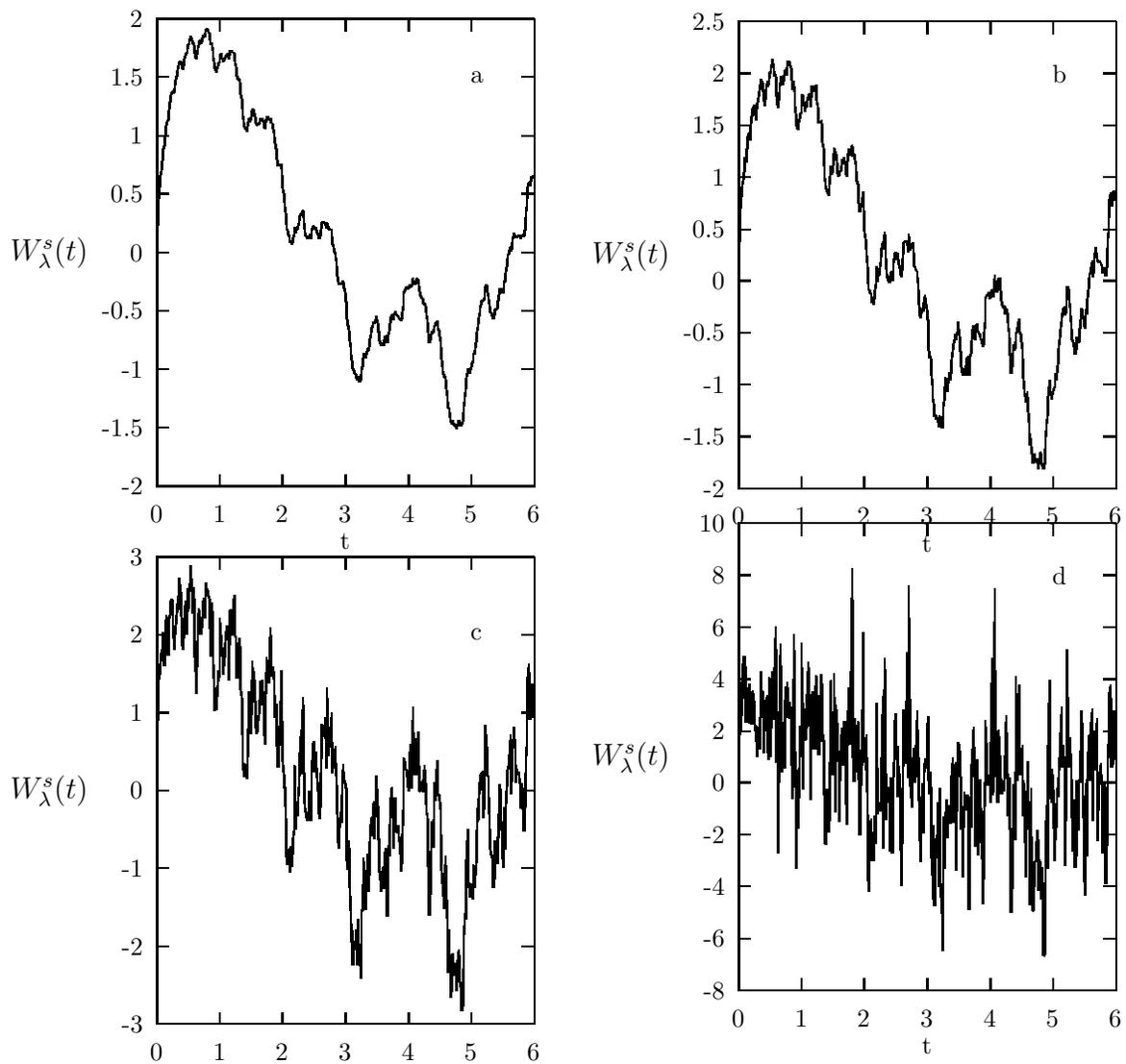

\begin{minipage}{430pt}
\input{isin5.tex}
\end{minipage} \\
\begin{minipage}{430pt}
\input{isin6.tex}
\end{minipage} \\ 
\begin{minipage}{430pt}
\input{isin7.tex}
\end{minipage} \\
\begin{minipage}{430pt}
\input{isin8.tex}
\end{minipage}
\caption{Graphs of Weierstrass functions for $\lambda = 1.5 $
and (a) $s=1.1$ (b) $s=1.3$ (c) $s=1.6$ (d) $s=1.9$.}\label{fg:wf}
\end{figure}

The graph of this function is known~\cite{3Fal} to be a fractal. 
Fractal dimension of
such and related curves were first studied in~\cite{3BU}.
The box dimension of the graph of this function is $s$~\cite{3Fal}. 
The Hausdorff
dimension of its graph is still unknown. The best known bounds are given
by Mauldin and Williams~\cite{3MW} where they have shown that there is
a constant $c$ such that
\begin{eqnarray}
s - {c \over{\log{\lambda}}} \leq \mbox{dim}_H\mbox{graph} f \leq s. \nonumber
\end{eqnarray}
Recently it was shown~\cite{3Hun} that if a random phase is added to 
each cosine term in (\ref{eq:Wgen}) then the Hausdorff dimension of the graph
of the resulting function is $2+\log{a}/\log{b}$.
These kind of functions have been studied in detail in~\cite{3Har,3BU,3BL,3Fal}.
Figure~\ref{fg:wf} shows graphs of the Weierstrass function for a fixed
value of $\lambda$ and various values of the  dimensions. It is clear that
as dimension increases the graph becomes more and more irregular. As the
value of the dimension approaches 2 the graph tends to fill the area.

\subsubsection{Weierstrass function as a limiting set of a dynamical system}
Nowhere differntiable functions arise as attractors or repellers of dynamical
systems. One such example was cited in chapter~\ref{ch:1}. Here we discuss
another example in detail.
Consider a mapping $h: I\!\!R^2 \rightarrow I\!\!R^2$ given by~\cite{3Fal}
\begin{eqnarray}
h(t,x)=(\lambda t, \lambda^{2-s}(x-\sin(t))), \label{eq:maph}
\end{eqnarray}
where $\lambda > 1$ and $0<s<2$.
Then we get 
\begin{eqnarray}
h^2(t,x)&=&h(\lambda t, \lambda^{2-s}(x-\sin(t))) \\
&=& (\lambda^2 t, \lambda^{2-s}( \lambda^{2-s}(x-\sin(t)) -\sin(\lambda t)))\\
&=& (\lambda^2 t, \lambda^{2(2-s)}(x - ( \sin(t)+ 
\lambda^{s-2} \sin(\lambda t)))
\end{eqnarray}
Therefore, continuing in the same fashion, we have the $n$th iterate as
\begin{eqnarray}
h^n(t,x)&=& (\lambda^n t, \lambda^{n(2-s)}(x - \sum_{k=1}^{n} 
{\lambda}^{(s-2)k} \sin{\lambda}^kt).
\end{eqnarray}
This shows that the graph of Weierstrass function~(\ref{eq:Wsp}) is a repeller
for $h$ in~(\ref{eq:maph}).

\section{Fractional differentiability of Weierstrass function} \label{se:2ch3}
Now we consider the Weierstrass function as a prototype example of a
continuous but nowhere differentiable function
(which is homogeneous in some sense to be explianed later)
and study its local fractional differentiability properties~\cite{3KG1}. 
We begin by considering the origin $t=0$. We first
find fractional derivative of $W_{\lambda}(t)$:
\begin{eqnarray}
{{d^qW_{\lambda}(t)}\over{dt^q}}  
&=& {\sum_{k=1}^{\infty} {\lambda}^{(s-2)k}{{d^q\sin({\lambda}^kt)}\over {dt^q}}}\nonumber\\
&=& {\sum_{k=1}^{\infty} {\lambda}^{(s-2+q)k}{{d^q\sin({\lambda}^kt)}\over {d({\lambda}^kt)^q}}}, \nonumber
\end{eqnarray}
provided the right hand side converges uniformly. Using the relation 
\begin{eqnarray}
{{d^q\sin(x)}\over {d x^q}}={{d^{q-1}\cos(x)}\over{d x^{q-1}}}, 
\;\;\;\;0<q<1,\nonumber
\end{eqnarray}
 we get
\begin{eqnarray}
{{d^qW_{\lambda}(t)}\over{dt^q}}  
&=& {\sum_{k=1}^{\infty} {\lambda}^{(s-2+q)k}{{d^{q-1}\cos({\lambda}^kt)}\over {d({\lambda}^kt)^{q-1}}}}.\label{a}
\end{eqnarray}
From second mean value theorem it follows that the fractional integral 
of $\cos({\lambda}^kt)$ of order $q-1$ is 
bounded uniformly for all values of ${\lambda}^kt$. 
This implies that the series on the right 
hand side will converge uniformly for $q<2-s$, justifying our taking
the fractional derivative operator inside the sum.

Also as $t \rightarrow 0$ for
any $k$ the fractional integral in the summation of equation (\ref{a}) goes to zero.
Therefore it is easy to see that
\begin{eqnarray}
I\!\!D^qW_{\lambda}(0) = {\lim_{t\rightarrow 0} {{d^qW_{\lambda}(t)}\over{dt^q}}}=0\;\;\;
{\rm for} \;\;\;q<2-s.
\end{eqnarray}
This shows that $q^{th}$ local derivative of the Weierstrass function exists and
is continuous, at $t=0$, for $q<2-s$.

To check the fractional differentiability at any other point, say $\tau$,
we use $t'=t-\tau$ and $\widetilde{W} (t' )=W(t'+\tau )-W(\tau)$ so that
$\widetilde{W}(0)=0$. We have
\begin{eqnarray}
\widetilde{W}_{\lambda} (t' ) &=& \sum_{k=1}^{\infty} {\lambda}^{(s-2)k} \sin{\lambda}^k(t' +\tau)-
\sum_{k=1}^{\infty} {\lambda}^{(s-2)k} \sin{\lambda}^k\tau \nonumber\\
&=&\sum_{k=1}^{\infty} {\lambda}^{(s-2)k}(\cos{\lambda}^k\tau \sin{\lambda}^kt' 
+
\sin{\lambda}^k\tau(\cos{\lambda}^kt' -1)) \label{c}
\end{eqnarray}
Taking fractional derivative of this with respect to $t'$ and following the 
same procedure we can show that the fractional derivative of the Weierstrass
function of order $q<2-s$ exists at all points.

For $q>2-s$,  right hand side of the equation (\ref{a})  seems to diverge
and our first step of taking fractional derivative operator inside the
sum itself may not be justified.
Therefor to prove that the LFD of order $q>2-s$ does not
exist we adopt a different approach.
We do this by showing that there exists a sequence of 
points approaching 0 along which
the limit of the fractional derivative of order $2-s < q <1$ does not exist.
We use the property of the Weierstrass function \cite{3Fal}, viz,
for each $t' \in [0,1]$ and $0 < \delta \leq {\delta}_0$
there exists $t$ such that $\vert t-t' \vert \leq \delta$ and
\begin{eqnarray}
c{\delta}^{\alpha} \leq \vert W(t)-W(t') \vert , \label{uholder}
\end{eqnarray}
where $c > 0$ and $\alpha=2-s$, provided $\lambda$ is sufficiently large.
 We consider the case of $t'=0$ and
$t>0$.
Define $g(t)=W(t)-ct^{\alpha}$.

The abovementioned property, alongwith  continuity, 
of the Weierstrass function guarantees the existence of a 
sequence of points $t_1>t_2>...>t_n>...\geq 0$ such that
$t_n \rightarrow 0$ as $n \rightarrow \infty$ and $g(t_n) = 0$ 
and $g(t)>0$ on $(t_n,\epsilon)$ for some $\epsilon>0$, for all
$n$ (It is not ruled out that $t_n$ may be zero for finite $n$). 
Let us define
\begin{eqnarray}
g_n(t)&=&0\;\;\;{\rm if}\;\;\;t\leq t_n \nonumber\\
&=&g(t)\;\;\; {\rm otherwise}.\nonumber
\end{eqnarray}
Therefore we have, for $0 <\alpha < q < 1$,
\begin{eqnarray}
{{d^qg_n(t)}\over{d(t-t_n)^q}}={1\over\Gamma(1-q)}{d\over{dt}}{\int_{t_n}^t{{g(y)}\over{(t-y)^{q}}}}dy,\nonumber
\end{eqnarray}
where $t_n \leq t \leq t_{n-1}$. We assume that left hand side of the above
equation exists.
Let
\begin{eqnarray}
h(t)={\int_{t_n}^t{{g(y)}\over{(t-y)^{q}}}}dy.\nonumber
\end{eqnarray}
Now $h(t_n)=0$ and $h(t_n+\epsilon)>0$, for a suitable $\epsilon$, as the integrand is positive.
Because of the continuity there must exist an ${\epsilon}'>0$ and 
${\epsilon}'<\epsilon $
such that $h(t)$ is increasing on $(t_n,{\epsilon}')$.
Therefore
\begin{eqnarray}
0 \leq {{d^qg_n(t)}\over{d(t-t_n)^q}} {\vert}_{t=t_n}\;\;\;\;n=1,2,3...  .
\end{eqnarray}
This implies that
\begin{eqnarray}
c\;{{d^qt^{\alpha}}\over{d(t-t_n)^q}} {\vert}_{t=t_n} \leq {{d^qW(t)}\over{d(t-t_n)^q}} {\vert}_{t=t_n}
\;\;\;\;n=1,2,3...
\end{eqnarray}
But we know from eqn(\ref{xp}) that, when $0<\alpha <q<1$,
the left hand side in the above inequality approaches infinity as $t\rightarrow 0$.
This implies that the right hand side of the above inequality does not
exist as $t \rightarrow 0$. This argument can be generalized 
for all nonzero $t'$ by
changing the variable $t''=t-t'$.
This concludes the proof.

Therefore the critical order of the Weierstrass function 
is $2-s$ at all points. Thus the maximum order of differentiability is
directly related to the dimensions.

\section{Critical order and L\'evy index of a L\'evy flight} 
Schlesinger et al \cite{3Shl} have considered a 
L\'evy flight on a one dimensional 
periodic lattice where a particle jumps from one lattice site 
to other  with the probability density given by
\begin{eqnarray}
P(x) = {{{\omega}-1}\over{2\omega}} \sum_{j=0}^{\infty}{\omega}^{-j}
[\delta(x, +b^j) + \delta(x, -b^j)]
\end{eqnarray}
where $x$ is magnitude of the jump, $b$ is a lattice spacing and $b>\omega>1$. 
$\delta(x,y)$ is a Kronecker delta.
The characteristic function for $P(x)$ is given by
\begin{eqnarray}
\tilde{P}(k) = {{{\omega}-1}\over{2\omega}} \sum_{j=0}^{\infty}{\omega}^{-j}
cos(b^jk),
\end{eqnarray}
which is nothing but the Weierstrass cosine function.
For this distribution the L\'evy index is $\log{\omega}/\log{b}$, which can be
identified as critical order of $\tilde{P}(k)$. 

More generally, for the L\'evy distribution with index $\mu$ 
the characteristic function
is given by
\begin{eqnarray}
\tilde{P}(k) =A \exp{c\vert k \vert^{\mu}}.
\end{eqnarray}
Critical order of this function at $k=0$
also turns out to be same as $\mu$. Thus the L\'evy index can be identified as
the critical order of the characteristic function at $k=0$.

\section{Equivalence between critical order and the H\"older exponent}
\label{se:equiv}
\begin{thm} \label{th:1}
 Let $f:[0,1]\rightarrow I\!\!R$ be a continuous function.

a)If
\begin{eqnarray}
\lim_{x\rightarrow y} {d^q(f(x)-f(y)) \over{[d(x-y)]^q}}=0\;\;\;
{\rm for}\;\; q<\alpha\;\; 
,\nonumber
\end{eqnarray}
where $q,\alpha \in (0,1)$,
for all $y \in (0,1)$, 
then  $dim_Bf(x) \leq 2-\alpha$.

b)If there exists a sequence  $x_n \rightarrow y$ as
$n \rightarrow \infty$ such that
\begin{eqnarray}
\lim_{n\rightarrow \infty} {d^q(f(x_n)-f(y)) \over{[d(x_n-y)]^q}}=\pm \infty\;\;\;
{\rm for}\;\; q>\alpha\;\; 
,\nonumber
\end{eqnarray}
for all $y$, 
then $dim_Bf \geq 2-\alpha$.
\end{thm}

\noindent
{\bf Proof}: See appendix A

Note that  part (a) of the theorem above is the generalization
of the statement that $C^1$ functions are locally Lipschitz (hence their
graphs have dimension 1) to the case when the function has H\"older type
upper bound (hence their dimension is greater than one).

Here the function is required to
have the same critical order throughout the interval. We can weaken this
condition slightly. Since we are dealing with box dimension which
is  finitely stable \cite{3Fal}, we can allow finite number of points having
different critical order so that we can divide the set in finite parts
having same critical order in each part.

We can also prove a partial converse of the above theorem.

\begin{thm} \label{th:2}

Let $f:[0,1]\rightarrow I\!\!R$ be a continuous function.

a) Suppose 
\begin{eqnarray}
\vert f(x)- f(y) \vert \leq c\vert x-y \vert ^{\alpha}, \nonumber
\end{eqnarray}
where $c>0$, $0<\alpha <1$ and $|x-y|< \delta$ for some $\delta >0$.
Then
\begin{eqnarray}
\lim_{x\rightarrow y} {d^q(f(x)-f(y)) \over{[d(x-y)]^q}}=0\;\;\;
{\rm for}\;\; q<\alpha\;\; 
\nonumber
\end{eqnarray}
for all $y\in (0,1)$

b) Suppose that for each $y\in (0,1)$ and for each $\delta >0$ there exists x such that
$|x-y| \leq \delta $ and
\begin{eqnarray}
\vert f(x)- f(y) \vert \geq c{\delta}^{\alpha}, \nonumber
\end{eqnarray}
where $c>0$, $\delta \leq {\delta}_0$ for some ${\delta}_0 >0$ and $0<\alpha<1$.
Then there exists a sequence $x_n \rightarrow y$ as $n\rightarrow \infty$
such that
\begin{eqnarray}
\lim_{n\rightarrow \infty} {d^q(f(x_n)-f(y)) \over{[d(x_n-y)]^q}}=\pm \infty\;\;\;
{\rm for}\;\; q>\alpha\;\; 
\nonumber
\end{eqnarray}
for all $y$.
\end{thm}

\noindent
{\bf Proof}

a)See appendix B

b)Proof follows by the method used in the previous section to show that
the fractional derivative of order greater than $2-\alpha$ of the Weierstrass
function does not exist.

These two theorems give an equivalence between H\"older exponent and the critical
order of fractional differentiability.

{
\appendix
\section{Appendix A: Proof of theorem 1}

 (a)Without loss of generality assume $y=0$ and $f(0)=0$.
We consider the case of $q<\alpha$.

As $0<q<1$ and $f(0)=0$ we can write \cite{3OS} 
\begin{eqnarray}
f(x)&=&{d^{-q}\over{d x^{-q}}}{d^qf(x)\over{d x^q}}\nonumber\\ 
&=&{1\over\Gamma(q)}{\int_0^x{{d^qf(y)\over{dy^q}}\over{(x-y)^{-q+1}}}}dy \label{comp}
\end{eqnarray}
Now
\begin{eqnarray}
\vert f(x)\vert \leq {1\over\Gamma(q)}{\int_0^x{\vert {d^qf(y)\over{dy^q}}\vert
\over{(x-y)^{-q+1}}}}dy \nonumber
\end{eqnarray}
As, by hypothesis, for $q<\alpha$,
\begin{eqnarray}
\lim_{x\rightarrow 0}{d^qf(x)\over{d x^q}}=0 \nonumber
\end{eqnarray}
 we have, for any $\epsilon > 0$, a $\delta > 0$ such that
$\vert {d^qf(x)\over{d x^q}}\vert < \epsilon$ for all $x< \delta$
\begin{eqnarray}
\vert f(x)\vert &\leq& {\epsilon\over\Gamma(q)}{\int_0^x{dy
\over{(x-y)^{-q+1}}}}\nonumber\\
&=&{\epsilon\over \Gamma(q+1)}x^q\nonumber
\end{eqnarray}
As a result we have
\begin{eqnarray}
\vert f(x)\vert &\leq& K \vert x\vert ^q \;\;\;\;{\rm for}\;\;\; x<\delta.\nonumber
\end{eqnarray}
Now this argument can be extended for general $y$ simply by considering
$x-y$ instead of $x$ and $f(x)-f(y)$ instead of $f(x)$. So finally
we get for $q<\alpha$
\begin{eqnarray}
\vert f(x)-f(y)\vert &\leq& K \vert x-y\vert ^q \;\;\;\;{\rm for}
\;\;\vert x-y \vert < \delta,\label{holder}
\end{eqnarray}
for all $y \in (0,1)$. Hence we have \cite{3Fal}
\begin{eqnarray}
\mbox{dim}_Bf(x) \leq 2-\alpha.\nonumber
\end{eqnarray}

b)Now we consider the case $q>\alpha$. If we have
\begin{equation}
\lim_{x_n\rightarrow 0}{d^qf(x_n)\over{dx_n^q}}=\infty \label{k0}
\end{equation}
then for given $M_1 >0$ and $\delta > 0$ we can find positive integer $N$ such that $|x_n|<\delta$ and
$ {d^qf(x_n)}/{dx_n^q} \geq M_1$ for all $n>N$. Therefore by eqn(\ref{comp})
\begin{eqnarray}
 f(x_n) &\geq& {M_1\over\Gamma(q)}{\int_0^{x_n}{dy
\over{(x_n-y)^{-q+1}}}}\nonumber\\
&=&{M_1\over \Gamma(q+1)}x_n^q\nonumber
\end{eqnarray}
If we choose $\delta=x_N$ then we can say that there exists $x<\delta$
such that
\begin{eqnarray}
f(x) \geq k_1 {\delta}^q \label{k1}
\end{eqnarray}
If we have
\begin{eqnarray}
\lim_{x_n\rightarrow 0}{d^qf(x_n)\over{dx_n^q}}=-\infty \nonumber
\end{eqnarray}
then for given $M_2 >0$ we can find positive integer $N$ such that
$ {d^qf(x_n)}/{dx_n^q} \leq -M_2$ for all $n>N$. Therefore
\begin{eqnarray}
 f(x_n) &\leq& {-M_2\over\Gamma(q)}{\int_0^{x_n}{dy
\over{(x_n-y)^{-q+1}}}}\nonumber\\
&=&{-M_2\over \Gamma(q+1)}x_n^q\nonumber
\end{eqnarray}
Again if we write $\delta=x_N$, there exists $x<\delta$ such that
\begin{eqnarray}
f(x) \leq  -k_2 {\delta}^q \label{k2}
\end{eqnarray}
Therefore by (\ref{k1}) and (\ref{k2})there exists $x<\delta$ such that, for $q>\alpha$, 
\begin{eqnarray}
\vert f(x)\vert &\geq& K \delta^q.\nonumber
\end{eqnarray}
Again for any $y \in (0,1)$ there exists $x$ such that
for $q>\alpha$ and $|x-y|<\delta$
\begin{eqnarray}
\vert f(x)-f(y)\vert &\geq& k \delta^q.\nonumber
\end{eqnarray}
 Hence we have \cite{3Fal}
\begin{eqnarray}
dim_Bf(x) \geq 2-\alpha.\nonumber
\end{eqnarray}

\section{Appendix B: Proof of the theorem 2a}
Assume that there exists a sequence  $x_n \rightarrow y$ as
$n \rightarrow \infty$ such that
\begin{eqnarray}
\lim_{n\rightarrow \infty} {d^q(f(x_n)-f(y)) \over{[d(x_n-y)]^q}}=\pm \infty\;\;\;
{\rm for}\;\; q<\alpha\;\; 
,\nonumber
\end{eqnarray}
for some $y$. Then by arguments following the equation~(\ref{k0}) 
 it is a 
contradiction.
Therefore
\begin{eqnarray}
\lim_{x\rightarrow y} {d^q(f(x)-f(y)) \over{[d(x-y)]^q}}=const\;\;
{\rm or}\;\; 0\;\;\;
{\rm for}\;\; q<\alpha\;\; 
\nonumber
\end{eqnarray}
Now if
\begin{eqnarray}
\lim_{x\rightarrow y} {d^q(f(x)-f(y)) \over{[d(x-y)]^q}}=const\;\;\;
{\rm for}\;\; q<\alpha,\;\; 
\nonumber
\end{eqnarray}
then we can write
\begin{eqnarray}
{d^q(f(x)-f(y)) \over{[d(x-y)]^q}}=K+\eta(x,y)\;\;\;
\nonumber
\end{eqnarray}
where $K=const$ and $\eta(x,y) \rightarrow 0$ 
sufficiently fast as $x\rightarrow y$. Now
taking $\epsilon$-derivative of both sides, for sufficiently small $\epsilon$ we get
\begin{eqnarray}
 {d^{q+\epsilon}(f(x)-f(y)) \over{[d(x-y)]^{q+\epsilon}}}={{K(x-y)^{-\epsilon}}
\over {\Gamma(1-\epsilon)}} + {d^{\epsilon}{\eta(x,y)}\over{[d(x-y)]^{\epsilon}}}
\;\;\;
{\rm for}\;\; q+\epsilon <\alpha\;\; 
\nonumber
\end{eqnarray}
As $x\rightarrow y$ the right hand side of the above equation goes
to infinity (term involving $\eta$ doesn't matter since $\eta$ goes to 0
sufficiently fast)
which again is a contradiction. Hence the proof.
}
\setcounter{chapter}{4}

\chapter{Pointwise Behavior of Irregular Functions} \label{ch:4}

\section{Motivation}
The relevance  of irregular functions in physical
processes was discussed  in Chapter~\ref{ch:0}.
There are also cases such as 
intermittent bursts  in velocity field of a turbulent
fluid, in chaotic signals etc. These and earlier examples demonstrate 
that the regularity of signals
in physical systems may vary from point to point, 
and call for an elaborate study of pointwise regularity of irregular signals. Multifractal functions
and measures are used to model these signals. 

There are several approaches to
study pointwise behavior of functions. Recently wavelet transforms 
~\cite{4Hol,4Jaf1}
have been used for this purpose and  have met with some success. 
In the following we demonstrate that LFD, as defined in chapter~\ref{ch:2},
is a tool that can be used to characterize
irregular functions and point out its certain advantages over 
 wavelet transforms.

Recall that 
H\"older exponent $\alpha(y)$ of a function $f$
at $y$ was defined in~\ref{se:holder} 
as the largest exponent such that there exists a polynomial
$P_n(x)$ of order $n$ that satisfies
\begin{eqnarray}
\vert f(x) - P_n(x-y) \vert = O(\vert x-y \vert^{\alpha}),
\end{eqnarray}
for $x$ in the neighborhood of $y$. Various authors \cite{4MBA,4Jaf2} have used this definition.
It is clear from theorem~\ref{th:1} that
LFDs provide a way to calculate H\"older exponents and 
dimensions. It may be noted that since there is a clear change
in behavior when order $q$ of the derivative crosses critical order
of the function 
it should be easy to determine the H\"older exponent numerically.
Earlier methods using autocorrelations for fractal signals \cite{4Fal} 
involve an additional step of estimating the autocorrelation itself.

We begin our discussion on use of LFD to study pointwise behavior by
considering isolated and masked singularities. Then we present a short
 introduction on the multifractal functions and study their 
local properties using LFD.

\section{Isolated singularities and masked singularities}
Let us start~\cite{4KG1,4KG2} with the case of isolated
singularities. We choose a simple example $f(x)=ax^{\alpha},\;\;\;0<\alpha
<1,\;\;\;x>0$, $a$ constant.  Critical order at $x=0$ gives the {\it order} of 
singularity at that point.

For the cases where two or more singularities are superposed 
on each other at the same
point,
 we can use LFD to detect a higher order singularity which may be
{\it masked} by 
one with lower order.
As demonstrated below, we can estimate and subtract the contribution due to 
lower order singularity from the 
function  and find out the critical order of the remaining function, 
in the following way.
We first find the critical order of the given function at the point of the
singularity. This itself is the order of the lower order singularity.
Then we determine the value of the LFD at the critical order. 
Consider, for example, the function
\begin{eqnarray}
f(x)=ax^{\alpha}+bx^{\beta},\;\;\;\;\;\;0<\alpha <\beta <1,\;\;\;x>0.
\label{masked}
\end{eqnarray}
LFD of this function at $x=0$ of the order $\alpha$ is 
$I\!\!D^{\alpha}f(0)=a\Gamma(\alpha+1)$.
Using this estimate of stronger (lower order) singularity we now write
\begin{eqnarray}
G(x;\alpha)=f(x)-f(0)-{I\!\!D^{\alpha}f(0)\over\Gamma(\alpha+1)}x^{\alpha}
\nonumber
\end{eqnarray}
which for the function $f$ in eqn(\ref{masked}) is
\begin{eqnarray}
{ {d^q G(x'\alpha) }
\over{d x^q}} = {b\Gamma(\beta+1)\over{\Gamma(\beta-q+1)}}x^{\beta-q}
\end{eqnarray}
Therefore the critical order of the  function $G$, at $x=0$, is $\beta$. 
 Notice that the estimation of the weaker (higher order) 
singularity was possible in the
above calculation just because the LFD gave the coefficient of $x^{\alpha}/
{\Gamma(\alpha+1)}$. This suggests that using LFD, one should be able to extract  singularity spectrum
masked by that of strong singularities. Hence one 
can gain more insight into the processes giving rise to irregular
behavior.

Comparison of two methods of studying pointwise behavior 
of functions, one using wavelets and other using LFD, 
shows that characterization of H\"older classes of
functions using LFD is direct and involves fewer assumptions. 

\begin{itemize}
\item It has 
been shown~\cite{4MBA} that by using wavelet transforms 
one can detect singularities masked by an $n$th order polynomial,
 by choosing analyzing wavelet with 
its first $n$ 
moments vanishing. If one has to extend the wavelet method 
for the unmasking of weaker singularities from lower order ones,
 one would then require analyzing wavelets with fractional moments vanishing.
If in addition there exists a polynomial masking both the singularities  
one will require the above condition along with the vanishing  
of first $n$ moments. 

\item Using wavelets to characterize
H\"older class of functions with oscillating singularity, 
e.g., $f(x)=x^{\alpha}sin(1/x^{\beta})$, $x>0$, $0< \alpha <1$ and 
$\beta>0$, 
will require two exponents \cite{4ABM}. 
On the other hand LFD of the function, through the theorems~\ref{th:1} and~\ref{th:2},  
directly  gives its H\"older exponent.

\item The class of functions to be analyzed using wavelet transforms is in
general restricted~\cite{4Jaf2}. These restrictions essentially arise
from the asymptotic properties of the wavelets used.
 On the other hand, with truly
local nature of LFD one doesn't have to bother about behavior of functions
outside our range of interest.

\end{itemize}

Now we consider the functions where we have singularities
of various orders at each and every point.

\section{Introduction to multifractal functions}
Extensive literature exists on the study of multifractal measures
and their applications
\cite{4BPPV,4HJKPS,4CLP,4JKP,4Man}. Their importance lies in 
 the fact that such measures are natural measures which are used in the
analyses of many physical phenomena \cite{4Fed,4Vic}. 
However there are cases in which the object
one wants to understand is a {\it function} (e.g. a fractal or multifractal signal)
rather than a set or a measure. For instance one would like to
characterize the velocity field of fully developed turbulence. 
Moreover, to every measure
we can assign a function (which is just its distribution function) but not vice
versa.

\begin{figure}
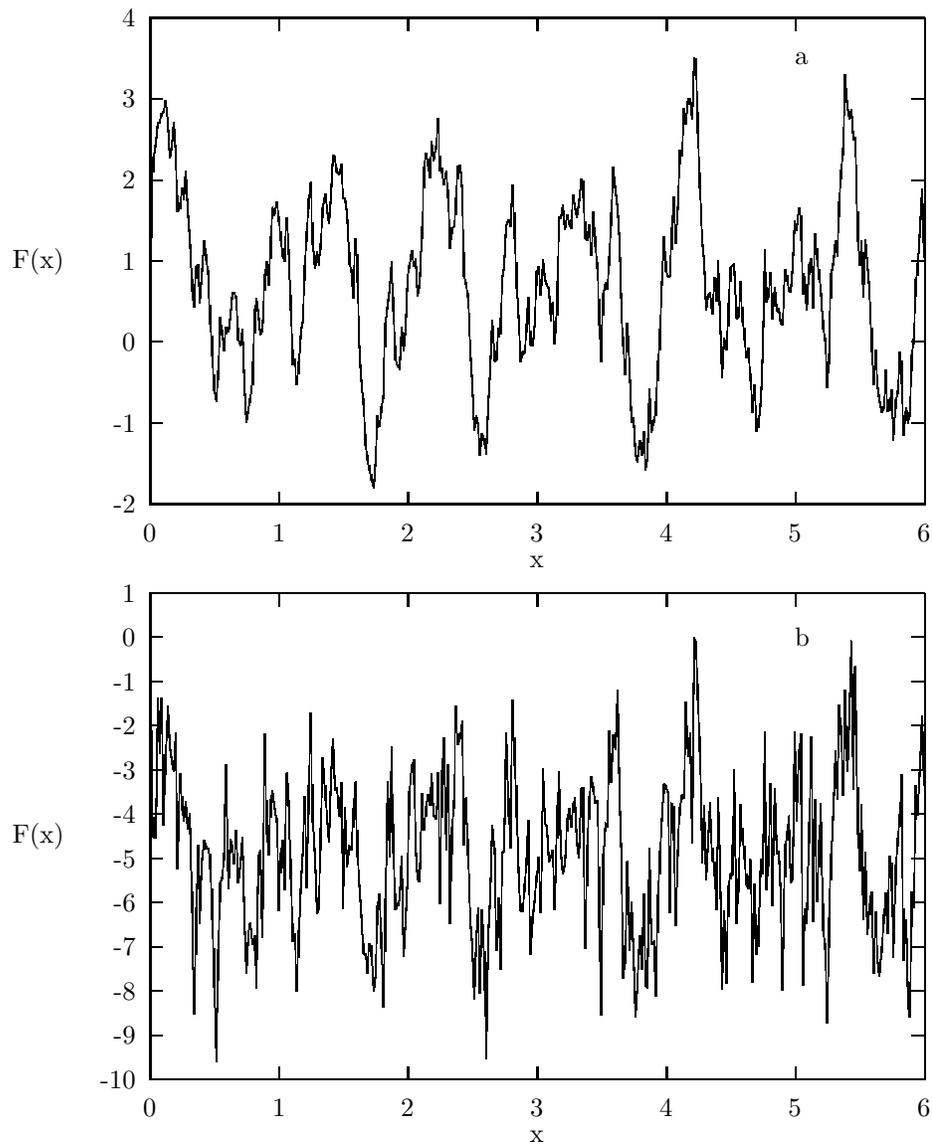
 
\input{iselfsim5.tex}
\input{iselfsim6.tex}
\caption{The graph of selfsimilar multifractal function $F(x)$ of 
equation~(\protect \ref{eq:mult}) with ${\mu}_1={\mu}_2=1/2$, $b_1=0$, $b_2=1/2$, and
(a) ${\lambda}_1=1.5^{-0.5}$, ${\lambda}_2=1.5^{-0.9}$
(b) ${\lambda}_1=1.5^{-0.5}$, ${\lambda}_2=1.5^{-0.1}$.
Please notice the intermittent bursts in the above functions.}\label{fg:mf}
\end{figure}

We saw in section \ref{se:1ch3} that the Weierstrass function 
is a fractal function, i.e., it has the same
H\"older exponent at every point. However, there exist more general functions
which do not have same H\"older exponent everywhere.
These are the abovementioned multifractal functions. 
Such functions can be used to model various intermittent phenomena -- spatial,
temporal as well as spatio--temporal -- arising
in physical systems. Such an intermittent behavior is clearly seen in
Fig.~\ref{fg:mf} which is a graph of a multifractal function.
Recall that in section~\ref{se:equiv}, while dealing with fractal functions, 
an equivalence between the H\"older
exponent and the critical order has been established.
The same equivalence also holds for multifractal functions where we have 
different
H\"older exponents at different points.
 Selfsimilar multifractal functions have been constructed by 
Jaffard~\cite{4Jaf2}.
Let us discuss one specific example of such a function.
This function is a solution $F(x)$ of the following functional equation 
\begin{eqnarray} 
F(x)=\sum_{i=1}^d {\lambda}_iF(S_i^{-1}(x)) + g(x),\;\;\;x\;\mbox{real}.
\label{eq:mult}
\end{eqnarray}
where $S_i$'s are the affine transformations of the kind 
$S_i(x)={\mu}_ix+b_i$ (with $\vert \mu_i \vert < 1$ and $b_i$'s real),
  $\lambda_i$'s 
are some real numbers, and $g$ is any sufficiently smooth function  
(it is assumed that $g$ and its
 derivatives  have fast decay). 
For the sake of illustration 
we choose the case $d=2$, with ${\mu}_1={\mu}_2=1/3$, $b_1=0$, $b_2=2/3$, 
${\lambda}_1=3^{-\alpha}$, ${\lambda}_2=3^{-\beta}$ ($0<\alpha<\beta<1$) and 
\begin{eqnarray}
g(x)&=& \sin(2\pi x)\;\;\;\;\;\;{\rm if}\;\;\;\; x\in [0,1]\nonumber\\
&=&0\;\;\;\;\;\;\;\;\;{\rm otherwise}. \nonumber
\end{eqnarray}
The graph of $F(x)$ with different values of parameters than above is plotted
in Fig.~\ref{fg:mf}.
Such functions are studied in detail in~\cite{4Jaf2} using wavelet transforms.
There it was shown that  the  above functional equation~(\ref{eq:mult}),
alongwith the parameters we have chosen above,
has a unique solution $F$. Furthermore at every point
$F$ either has H\"older exponents ranging from 
$\alpha$ to $\beta$ or is as smooth as $g$.  A sequence of points $S_{i_1}(0),\;\;$$ \;S_{i_2}S_{i_1}(0),\;\;$ $ 
\cdots,\;\;\; $$S_{i_n}\cdotp \cdotp \cdotp $$S_{i_1}(0), \;\;\cdots$,
where $i_k$ takes values 1 or 2, 
tends to a point in $[0,1]$ (in fact to a point of a triadic
Cantor set) and for the values of 
${\mu}_i$s chosen above this correspondence between sequences and limits 
is one to one. 
The solution of the above functional equation is given by~\cite{4Jaf2} 
\begin{eqnarray}
F(x)=\sum_{n=0}^{\infty}\;\;\;\sum_{i_1,\cdots,i_n=1}^2{\lambda}_{i_1}\cdots{\lambda}_{i_n}
g(S_{i_n}^{-1}\cdots S_{i_1}^{-1}(x)). \label{soln}
\end{eqnarray}
Note that with the above choice of parameters the inner sum in (\ref{soln})
 reduces to a single term. Jaffard~\cite{4Jaf2} has shown that the local 
H\"older exponent at $y\in [0,1]$ is
\begin{eqnarray}
h(y)=\liminf_{n\rightarrow \infty}{{\log{({\lambda}_{{i_1}(y)}\cdots{\lambda}_{{i_n}(y)})}}
\over{\log{({\mu}_{{i_1}(y)}\cdots{\mu}_{{i_n}(y)})}}},
\end{eqnarray}
where $\{i_1(y),\cdot\cdot\cdot, i_n(y)\}$ is a sequence of integers 
depending on a point $y$ (i.e. $S_{i_1(y)} \cdot \cdot \cdot 
S_{i_n(y)} \rightarrow y$).
 It is clear that $h_{min}=\alpha$ and
$h_{max}=\beta$. The function $F$ at the points of a triadic cantor
 set has $h(x) \in [\alpha , \beta]$
and at other points it is smooth (as smooth as $g$).
 Benzi et. al.~\cite{4Ben} have constructed multifractal functions which
are random in nature unlike the above nonrandom functions.
For still another approach see~\cite{4Jun}.

Several well-known pathological functions have been reanalyzed 
in~\cite{4Jaf3,4JM,4DL,4Jaf4} 
and found to have multifractal nature.
Here we consider one classic example of multifractal function.
\begin{eqnarray}
R(x) = \sum_{n=1}^{\infty}{1\over{n^2}}\sin(\pi n^2x)
\end{eqnarray}
This function was proposed by Riemann. It turns out that the regularity
of this function varies strongly from point to point. Hardy and Littlewood
\cite{4HL} proved that $R(x)$ is not differentiable at $x_0$ if $x_0$
is irrational or if $x_0$ can not be written in the form 
$2p+1/2q+1$ ($p,q \in N$).
In fact they showed that the H\"older exponent at these points is less than
or equal to 3/4.
Gerver~\cite{4Ger} proved the differentiability of $R(x)$ at points
which can be written as
$2p+1/2q+1$ ($p,q \in N$). At these points the exponent is 3/2.
This function has also been studied in~\cite{4HT,4Dui}.
Jaffard \cite{4Jaf3} has recently shown that the dimension 
spectrum of $R(x)$ is
\begin{eqnarray}
d(\alpha) = \left\{ 
\begin{array}{ll}
4\alpha -2 & \mbox{if}\;\;\; \alpha \in [{1\over2},{3\over4}]\\
0 & \mbox{if}\;\;\; \alpha = {3\over2}\\
-\infty & \mbox{otherwise}, 
\end{array}   \right.
\end{eqnarray}
where $d(\alpha)$ is the Hausdorff dimension of the set where the 
H\"older exponent is $\alpha$.

\section{Treatment of multifractal function}

 We now proceed
with the analysis of multifractal functions using LFD~\cite{4KG1}. 
 Since LFD gives the local 
 behavior of the function, theorem~\ref{th:1} can also
be applied to  multifractal functions.
Jaffard's construction of a selfsimilar multifractal function has been studied
in the last section.
We consider the example~(\ref{soln}) and study its LFD: 
\begin{eqnarray}
{d^q(F(x)-F(y))\over{[d(x-y)]^q}}&=&\sum_{n=0}^{\infty}\;\;\;\sum_{i_1,\cdots,i_n=1}^2{\lambda}_{i_1}\cdots{\lambda}_{i_n}\nonumber\\
&&\;\;\;\;\;\;\;\;\;{d^q[g(S_{i_n}^{-1}\cdots S_{i_1}^{-1}(x))-g(S_{i_n}^{-1}\cdots S_{i_1}^{-1}(y))]
\over{[d(x-y)]^q}}\nonumber\\
&=&\sum_{n=0}^{\infty}\;\;\;\sum_{i_1,\cdots,i_n=1}^2{\lambda}_{i_1}\cdots{\lambda}_{i_n}
({\mu}_{i_1}\cdots{\mu}_{i_n})^{-q} \nonumber\\
&&\;\;\;\;\;\;\;\;\;\;{d^q[g(S_{i_n}^{-1}\cdots S_{i_1}^{-1}(x))-g(S_{i_n}^{-1}\cdots S_{i_1}^{-1}(y))]
\over{[d(S_{i_n}^{-1}\cdots S_{i_1}^{-1}(x-y))]^q}}, \label{fdj}
\end{eqnarray}
provided the RHS is uniformly bounded.
Following the procedure described in section~\ref{se:2ch3} the fractional
derivative in the RHS can easily be seen to be uniformly bounded and 
series is convergent if $q<\min\{h(x),h(y)\}$.
Further it vanishes in the limit $x\rightarrow y$. Therefore if $q<h(y)$,
 $I\!\!D^qF(y)=0$, as
 in the case of Weierstrass function, showing that $h(y)$ is lower
bound on critical order. 

The procedure of finding an upper bound is lengthy.
It is relegated to the Appendix below where it is shown that upper bound
is also $h(y)$.

In this way an intricate analysis of finding out the lower bound on
H\"older exponent has been replaced by a calculation involving few steps. This
calculation can easily be extended to cases
incorporating more general  functions $g(x)$. 

Summarizing, the LFD enables one to calculate the local H\"older exponent even
for the case of multifractal functions. This fact, proved in 
theorems~\ref{th:1} and~\ref{th:2},
has been demonstrated with concrete illustration.

{
\appendix
\section{Appendix}
Our aim in this is appendix is to
 find an upper bound on the critical order and hence on the local
H\"older exponent for the function of equation (\ref{soln}). Our procedure will be similar to that of Jaffard \cite{4Jaf2} the only difference 
being that we  take fractional derivative instead of wavelet transform.
We proceed as follows. The defining equation for $F(x)$ when iterated 
$N$ times gives
\begin{eqnarray}
F(x)&=&\sum_{n=0}^{N-1}\;\;\;
\sum_{i_1,\cdots,i_n=1}^2{\lambda}_{i_1}\cdots{\lambda}_{i_n}
g(S_{i_n}^{-1}\cdots S_{i_1}^{-1}(x))\nonumber\\
&+& \sum_{i_1,\cdots,i_N=1}^2{\lambda}_{i_1}\cdots{\lambda}_{i_N}
F(S_{i_N}^{-1}\cdots S_{i_1}^{-1}(x))\;\;\;\;\;\;x\in [0,1].
\end{eqnarray}
We now consider
\begin{eqnarray}
{d^q(F(x)-F(y))\over{[d(x-y)]^q}}&=&\sum_{n=0}^{N-1}\;\;\;\sum_{i_1,\cdots,i_n=1}^2{\lambda}_{i_1}\cdots{\lambda}_{i_n}
({\mu}_{i_1}\cdots{\mu}_{i_n})^{-q} \nonumber\\
&&\;\;\;\;\;\;\;\;\;\;{d^q[g(S_{i_n}^{-1}\cdots S_{i_1}^{-1}(x))-g(S_{i_n}^{-1}\cdots S_{i_1}^{-1}(y))]
\over{[d(S_{i_n}^{-1}\cdots S_{i_1}^{-1}(x-y))]^q}}\nonumber\\
&&+\sum_{i_1,\cdots,i_N=1}^2{\lambda}_{i_1}\cdots{\lambda}_{i_N}
({\mu}_{i_1}\cdots{\mu}_{i_N})^{-q} \nonumber \\
&&\;\;\;\;\;\;\;\;\;\;{d^q[F(S_{i_N}^{-1}\cdots S_{i_1}^{-1}(x))-F(S_{i_N}^{-1}\cdots S_{i_1}^{-1}(y))]
\over{[d(S_{i_N}^{-1}\cdots S_{i_1}^{-1}(x-y))]^q}}. \label{iterated}
\end{eqnarray}
Let us denote the first term on RHS by $A$ and the second term by $B$.
In the following $y\in (0,1)$.
Choose $N$ such that $3^{-(N+1)}<\vert x-y \vert < 3^{-N}$.
 Denote $\lambda_{n(y)}={\lambda}_{{i_1}(y)}\cdots{\lambda}_{{i_n}(y)}$.
For the values of $\mu_i$s we have chosen ${\mu}_{i_1}\cdots{\mu}_{i_n}
= 3^{-n}$.
Now since $g$ is smooth $\vert g(x)-g(y) \vert \leq C \vert x-y \vert$.
\begin{eqnarray}
{d^q[g(S_{i_n}^{-1}\cdots S_{i_1}^{-1}(x))-g(S_{i_n}^{-1}\cdots S_{i_1}^{-1}(y))]
\over{[d(S_{i_n}^{-1}\cdots S_{i_1}^{-1}(x-y))]^q}}
& \leq & C \vert {{x-y}\over{{\mu}_{i_1}\cdots{\mu}_{i_n}}} \vert^{1-q}
\end{eqnarray}
From the way we have chosen $\mu_i$s and $\vert x-y \vert$ 
this term is bounded by
$C 3^{n(1-q)} 3^{-N(1-q)}$
Therefore the first term in equation (\ref{iterated}) above is bounded by
\begin{eqnarray}
 A &\leq & C\sum_{n=0}^{N-1} \lambda_{n(y)} 3^{nq} 
 3^{n(1-q)} 3^{-N(1-q)}  \nonumber\\
&=& C3^{-N(1-q)}\sum_{n=0}^{N-1} \lambda_{n(y)} 3^{n} \nonumber\\
&\leq & C3^{-N(1-q)}\lambda_{(N-1)(y)} 3^{N-1} (1+{\lambda_{(N-2)(y)}\over{
3\lambda_{(N-1)(y)}}}+{\lambda_{(N-3)(y)}\over{3^2\lambda_{(N-1)(y)}}}+\cdots)
\end{eqnarray}
The quantity in the bracket is bounded. Therefore
\begin{eqnarray}
 A &\leq & C3^{Nq}\lambda_{N(y)}  
\end{eqnarray}
Now we consider the second term in the RHS of equation (\ref{iterated}) and find a
lower bound on that term.
We assume that $F(x)$ is not Lipschitz at $y$ (for otherwise LHS of
equation (\ref{iterated}) will be zero in the limit $x\rightarrow y$ and the 
case is uninteresting).
Therefore there exists a sequence of points $x_n$ approaching $y$ such that
\begin{eqnarray}
\vert F(x_n)-F(y) \vert \geq c \vert x_n - y \vert \label{nonsmooth}
\end{eqnarray} 
The function is not Lipschitz at $y$ implies that it not Lipschitz at
$S_{i_N}^{-1}\cdots S_{i_1}^{-1}(y)$. Therefore we get (one may recall
that only one of the $2^n$ terms in the summation 
$\sum_{i_1,\cdots,i_N=1}^2$ is nonzero)
\begin{eqnarray}
B&\geq & c\lambda_{N(y)} 3^{Nq} 
 3^{N(1-q)} 3^{-N(1-q)} \nonumber\\
&\geq & c \lambda_{N(y)} 3^{Nq}
\end{eqnarray}
Therefore there exists a sequence $x_n$ such that
\begin{eqnarray}
\vert {d^q(F(x_n)-F(y))\over{[d(x_n-y)]^q}}-c \lambda_{N(y)} 3^{Nq} \vert
\leq  C \lambda_{N(y)} 3^{Nq}
\end{eqnarray}
Since (\ref{nonsmooth}) is valid for every $c$ for large enough $n$
$${d^q(F(x_n)-F(y))\over{[d(x_n-y)]^q}}\geq C' \lambda_{N(y)} 3^{Nq}$$
where $C'=c-C$
Now from the expression of $h(y)$ it is clear that when $q>h(y)$ 
LFD does not exist and the critical order is bounded from above also by $h(y)$.
}

\setcounter{chapter}{5}

\chapter{Local Fractional Differential Equations} \label{ch:5}

\section{Motivation and Introduction}
The most important equations of physical sciences are 
differential or integral equations, two themes emerging 
from the seventeenth-century
calculus. The derivatives of integer order are coefficients of
integer powers in the Taylor expansion. In particular the first
order derivative models a local linear behavior and the existence of
the first order derivative of the function implies Lipschitz continuity
(i.e., the H\"older exponent $\geq$ 1 as defined in~\ref{se:holder})
of the function. Therefore it is not possible to have H\"older
continuous functions as solutions to ordinary differential equations.
Such functions arise while dealing with phenomena involving fractals.
Hence it becomes necessary to develop calculus which would go beyond
ordinary calculus and incorporate H\"older continuos functions too.

The local fractional derivatives that we have studied 
in previous chapters preserve the local nature of the derivative operator.
Moreover they characterize the H\"older exponent of an irregular function.
This suggests these LFDs may provide a much needed tool for 
doing calculus for fractal
space-time. Hence it becomes pertinent to ask the questions such as:
what is the inverse of the LFD operator (if it exists), 
or can one write and solve
 equations involving LFDs? Also, what meaning and applications these
local fractional differential equations will potentially have?
These questions open a big area.
We'll try to answer some of these questions in the following sections.

\section{Local fractional differential equation}
In this section we  introduce a new type of equation which involve LFDs. 
We call them {\it local 
fractional
differential equations} (LFDE). 
These are equations involving LFD of a dependent variable with respect to
independent variables, as in the case of ordinary differential equations.
In order to understand the meaning of such equations
let us consider a simple example of the 
LFDE:
\begin{eqnarray}
I\!\!D^q_xf(x) = g(x). \label{eq:slfde}
\end{eqnarray}
In the above equation $g(x)$ is a known function and we have to find $f(x)$
such that this equation is satisfied.
The issue of the conditions guaranteeing solutions of such an 
equation is too involved to be addressed here in its full generality.
Therefore we restrict ourselves to specific cases.

\begin{figure}
\input{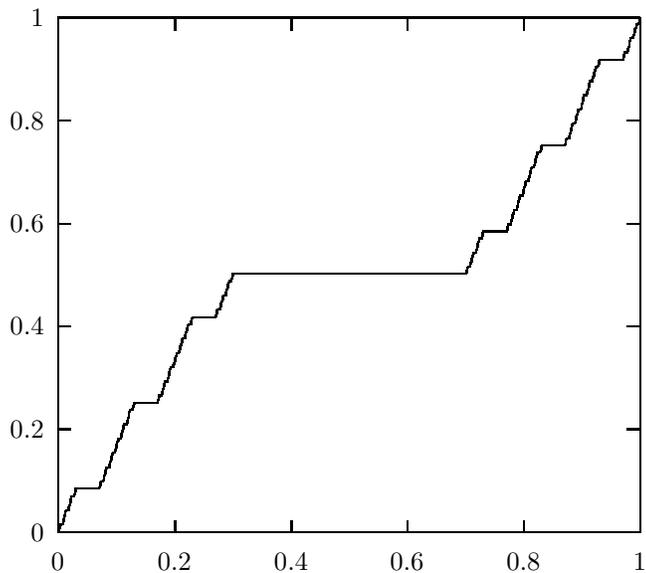}
\caption{A Lebesgue-Cantor staircase function}\label{fg:lcsf}
\end{figure}

We begin by noting
that the equation 
\begin{eqnarray}
I\!\!D^q_xf(x)= \mbox{Const},
\end{eqnarray}
 does not have
a finite solution when $0<q<1$. Interestingly, the solutions to (\ref{eq:slfde})
can exist, when $g(x)$ has a fractal support. For instance, when 
$g(x)=\chi_C(x)$, the membership function of a cantor set $C$
(i.e. $g(x)=1$ if $x$ is in $C$ and $g(x)=0$ otherwise), the solution with
initial condition $f(0)=0$ exists if $q=\alpha (\equiv \mbox{dim}_HC)$.
Explicitly, generalizing the Riemann integration procedure, 
\begin{eqnarray}
f(x)
=\lim_{N\rightarrow \infty} \sum_{i=0}^{N-1} {(x_{i+1}-x_i)^\alpha
\over\Gamma(\alpha+1)} F_C^i
\equiv { P_C(x)\over\Gamma(\alpha+1)}, \label{eq:PCt}
\end{eqnarray}
where $x_i$ are  subdivision points of the interval $[x_0=0, x_N=x]$ and
$F_C^i$ is a flag function which takes value 1 if the interval $[x_i,x_{i+1}]$
contains a point of the set $C$ and 0 otherwise. 
Now if we devide the interval $[0,x]$ into equal subintervals and denote
$\Delta=\Delta_i = x_{i+1}-x_i$ then we have $P_C(x)= \Delta^\alpha \sum F_C^i$.
But $\sum F_C^i$ is of the order of  $N^{-\alpha}$. Therefore
$P_C(x)$ satisfies the bounds 
$ax^\alpha \leq P_C(x) \leq bx^\alpha$ where $a$ and  $b$
are suitable positive constants. 
Note that $P_C(x)$ is a Lebesgue-Cantor 
(staircase) function as shown in the Fig.~\ref{fg:lcsf}. Since the function $\chi_C(x)$
is zero almost everywhere the function $P_C(x)$ is constant almost everywhere.
As is clear from the equation~(\ref{eq:PCt}), it rises only at points where $\chi_C(x)$ is non-zero.
In general, the algorithm of the equation
(\ref{eq:PCt}) is expected to work only for the sets $C$ for which 
$\mbox{dim}_BC=\mbox{dim}_HC$ (in fact in this case only $N^\alpha$
terms in the summation are nonzero).
It can now be seen that if $g(x)=\mbox{Const}$ then the sum in 
equation~(\ref{eq:PCt}) does not exist.
Therefore the equation $I\!\!D^q_xf(x)= \mbox{Const}$ does not have a solution
as stated earlier.
Though this algorithm possibly can be extended for more general sets we choose
to restrict ourselves to the sets satisfying the above condition.

It may be noted that the above LFDE is
different from usual fractional differential equations.
Conventional fractional differential equations, i.e., equations involving
Riemann-Liouville fractional derivatives, have been studied 
in the literature. These equations have found applications at various
places, e.g., solution of Bessel's equation, diffusion on curved surface,
etc \cite{5OS}. In fact the equations appearing in 
\cite{5Non,5GN,5GR,5RG,5Wys,5SW,5Jum,5Zas} are of this type. 
Since the Riemann-Liouville fractional 
derivative can be written as a integral of some function these equations
are essentially integro-differential equations.
On the other hand the equation we have proposed is intrinsically local.  
As mentioned in chapter~\ref{ch:2}, LFD $I\!\!D^qf(y)$ gives the
coefficient of the power $(x-y)^q$ in the fractional Taylor expansion
and hence local fractional differential
equations are equations satisfied by these coefficients. We feel that this is
a more natural generalization of ordinary differential equations. This is
not only due to the fact that the local nature is preserved but also because
the ordinary differential equations can  be interpreted in similar manner,
i.e., they are equations among the coefficients of integer powers $(x-y)^n$
in the expansion of a function $f(x)$ at $y$.

The equation~(\ref{eq:slfde}), along with its solution~(\ref{eq:PCt})
for the special choice of $g(x)$, has
an interesting consequence. When $q=\alpha=1$, i.e. the set $C$ has dimension 1,
we get a equation $f'(x)= \chi_C(x)$. The solution $f(x)$ of this equation gives
the length of the set $[0,x]\bigcap C$. In the similar manner,
the solution of the equation~(\ref{eq:slfde}) with $g(x)=\chi_C(x)$
would  yield
the fractal measure of the set $[0,x]\bigcap C$. This indicates 
that the operation
of the inversion of the LFD, a `local fractional integral'
(not to be confused with fractional integral), could yield 
a definition of the fractal measure.
It is interesting to note that this definition,
unlike other definitions of fractal measures, would have 
its origin in calculus. This
definition is equivalent, atleast in the case of simple sets considered here,
to the Hausdorff measure except for the normalization constant. 
The normalization constant $1/\Gamma(\alpha+1)$ in equation~(\ref{eq:PCt})
comes from
calculus whereas the one in Hausdorff measure (equation~(\ref{eq:norm}))
 comes from the geometry
of the covering elements. Of course, 
all these observations regarding the equivalence between two 
approaches, other than the normalization constant, remains to be rigorously
proved for general sets. But this interesting observation makes it plausible
to answer the questions raised in the chapter \ref{ch:0} about the relation
between `$\alpha$-fold' integral and the $\alpha$-dimensional measure.

\section{Derivation of local fractional Fokker-Planck equation}
The basic idea of this section is to utilize the fractional
Taylor expansions, developed in chapter~\ref{ch:2}, in
the Chapman-Kolmogorov condition and obtain analogs of 
Fokker-Planck-Kolmogorov (FPK) equation in the LFD calculus formalism.
We briefly recall the usual procedure of deriving the
FPK equation~\cite{5Ris}.
Let $W(x,t)$ denote the probability density for a random variable
$X$ taking value $x$ at time $t$. Then the Chapman-Kolmogorov equation is
\begin{eqnarray}
W(x,t+\tau)=\int P(x,t+\tau | x',t) W(x',t) dx', \label{CK}
\end{eqnarray}
where  $P(x_1,t_1 | x_2,t_2)$
denotes the transition probability from $x_2$ at time 
$t_2$ to $x_1$ at time $t_1$
and $\tau \geq 0$. 
In order to obtain the usual FPK equation~\cite{5Ris} we first expand
the integrand in the
equation~(\ref{CK}) in a Taylor series with respect to $x-x'$.
Further we assume that the moments,
\begin{eqnarray}
M_n(x,t,\tau)=\int dy (y-x)^n P(y,t+\tau|x,t),
\end{eqnarray}
can also be expanded into Taylor series with respect to $\tau$ as
\begin{eqnarray}
{M_n(x,t,\tau)\over{n!}} = A^n(x,t) \tau + O(\tau^2). \label{eq:intmnt}
\end{eqnarray}
Then various powers of $\tau$ are equated by expanding LHS with respect
to $\tau$.
As a result one arrives at the FPK equation
\begin{eqnarray}
{\partial W(x,t)\over{\partial t}} = {\cal{L}}(x,t)W(x,t), \label{eq:FPK}
\end{eqnarray}
where the Fokker-Planck operator ${\cal{L}}(x,t)$ is 
\begin{eqnarray}
{\cal{L}}(x,t)&\equiv&-{\partial\over{\partial x}}
A^1(x,t) +{\partial^2\over{\partial x^2}}
A^2(x,t). \label{eq:FPKop}
\end{eqnarray}

There are a number of limitations of this approach because of various 
assumptions going into its derivation. For instance, 
as noted in \cite{5LM}, probability
distributions whose second moment does not exist are not described by FPK 
equation even though such distributions may satisfy original 
Chapman-Kolmogorov equation.
Also, as emphasized in \cite{5Fel2}, the differentiability assumption going
into the Taylor expansion of the integrand of equation~(\ref{CK}) may also 
break down in various situations. For instance, the transitional probability
density may not be differentiable at $x=x'$; in which case the derivation
of FPK equation itself will break down. Another situation arises when
we have a fractal (nondifferentiable)
function as the initial probability density. In such a case
even the usual Fokker-Planck operator can not be operated on the initial 
density. 

It is thus of interest to broaden the class of differential  
equations one can derive starting from the Chapman-Kolmogorov equation, 
and study various 
processes described by them. In this section we 
investigate the possibility of dealing with those probability densities
which need not satisfy usual differentiability criteria.
We follow the standard method to derive the  
Fokker-Planck equation from equation~(\ref{CK}) outlined above with the 
modification that we now expand
the integrand using
fractional Taylor expansion~(\ref{taylor2}) instead of 
ordinary Taylor expansion. Thus, if $\Delta = x-x'$,
we have to consider two cases $\Delta \geq 0$ and $\Delta \leq 0$ separately. 
When $\Delta \geq 0$
\begin{eqnarray}
P(x,t+\tau | x',t) W(x',t) &=& P(x-\Delta +\Delta,t+\tau | x-\Delta,t) W(x-\Delta,t)\\
&=&\sum_{n=0}^N {1\over\Gamma(n+1)}\Delta^n({\partial\over{\partial(-x)}})^n
P(x+\Delta,t+\tau | x,t) W(x,t)\nonumber \\
&&+ {1\over\Gamma(\beta+1)}{\Delta}^{\beta}
I\!\!D_-^{\beta}[P(x+\Delta,t+\tau | x,t) W(x,t)],
\end{eqnarray}
and when $\Delta \leq 0$
\begin{eqnarray}
P(x,t+\tau | x',t) W(x',t) &=& P(x-\Delta +\Delta,t+\tau | x-\Delta,t) W(x-\Delta,t)\\
&=&\sum_{n=0}^N {1\over\Gamma(n+1)}(-\Delta)^n({\partial\over{\partial(-x)}})^n
P(x+\Delta,t+\tau | x,t) W(x,t)\nonumber \\
&&+ {1\over\Gamma(\beta+1)}{(-\Delta)}^{\beta}
I\!\!D_+^{\beta}[P(x+\Delta,t+\tau | x,t) W(x,t)],
\end{eqnarray}
where $N<\beta\leq N+1$.
This is the only place where our derivation differs from that of conventional
Fokker-Plank equation. (In the conventional derivation one expands the
integrand in equation (\ref{CK}) in Taylor series 
assuming implicitly that the probability 
densities involved are analytic.)
Substituting this into equation (\ref{CK}) we get
\begin{eqnarray}
W(x,t+\tau)&=&W(x,t)\! + \!\sum_{n=1}^N {1\over\Gamma(n+1)}({\partial\over{\partial(-x)}})^n
\int dx' \Delta^n P(x+\Delta,t+\tau | x,t) W(x,t)\nonumber \\
&&+ {1\over\Gamma(\beta+1)}
I\!\!D_{x-}^{\beta}[\int_x^{\infty} dy {(y-x)}^{\beta} 
P(y,t+\tau | x,t) W(x,t)]\nonumber\\
&&+ {1\over\Gamma(\beta+1)}
I\!\!D_{x+}^{\beta}[\int_{-\infty}^x \!\! dy {(x-y)}^{\beta} 
P(y,t+\tau | x,t) W(x,t)]+ \! Remainder \nonumber
\end{eqnarray}
where $I\!\!D_x$ is a partial LFD w.r.t $x$.
Now if $0< \alpha \leq 1$
\begin{eqnarray}
W(x,t+\tau)-W(x,t) = {\tau^{\alpha}I\!\!D_t^{\alpha}W(x,t)\over\Gamma(\alpha+1)}
+ Remainder. \nonumber
\end{eqnarray}
where $I\!\!D_t$ is partial LFD w.r.t. $t$.
In general $\alpha$ and $\beta$ may depend on $x$ and $t$.  But we 
assume that $\alpha$ and $\beta$ are constants.
Therefore we get
\begin{eqnarray}
{\tau^{\alpha}I\!\!D_t^{\alpha}W(x,t)\over\Gamma(\alpha+1)} &=&
\sum_{n=1}^N \big({\partial\over{\partial(-x)}}\big)^n
\big[{M_n(x,t,\tau)\over\Gamma(n+1)}W(x,t)\big] \nonumber \\
&&+ 
I\!\!D_{x-}^{\beta}\big[{M_{\beta}^+(x,t,\tau)\over\Gamma(\beta+1)}W(x,t)\big]
\nonumber \\ 
&& + 
I\!\!D_{x+}^{\beta}\big[{M_{\beta}^-(x,t,\tau)\over\Gamma(\beta+1)} W(x,t)\big]
\end{eqnarray}
where
\begin{eqnarray}
M_a^+(x,t,\tau)=\int_x^{\infty} dy (y-x)^a P(y,t+\tau|x,t)\;\;\; \;a>0,
\end{eqnarray}
\begin{eqnarray}
M_a^-(x,t,\tau)=\int_{-\infty}^x dy (x-y)^a P(y,t+\tau|x,t) \;\;\;\;a>0
\end{eqnarray}
and
\begin{eqnarray}
M_a(x,t,\tau)=M_a^+(x,t,\tau)+M_a^-(x,t,\tau) \label{eq:moments}
\end{eqnarray}
are transitional moments.
The limit $\tau \rightarrow 0$ gives us the
{\bf local fractional Fokker-Planck-Kolmogorov} (FFPK) equation 
\begin{eqnarray}
I\!\!D_t^{\alpha}W(x,t) = {\cal{L}_\alpha^\beta}(x,t)W(x,t) \label{trunKM}
\end{eqnarray}
where the operator $\cal{L}_\alpha^\beta$ is given by
\begin{eqnarray}
{\cal{L}_\alpha^\beta}(x,t)\equiv&& 
\sum_{n=1}^N \big({\partial\over{\partial(-x)}}\big)^n
A_{\alpha}^n(x,t)  +
I\!\!D_{x-}^{\beta}A_{\alpha -}^{\beta}(x,t) +
I\!\!D_{x+}^{\beta}A_{\alpha +}^{\beta}(x,t)  
\end{eqnarray}
where
\begin{eqnarray}
A_{\alpha \mp}^{\beta}(x,t)\equiv \lim_{\tau \rightarrow 0}
{M_{\beta}^\pm(x,t,\tau)\Gamma(\alpha+1)\over{\tau^{\alpha}\Gamma(\beta+1)}}
\label{mbeta}
\end{eqnarray}
and
\begin{eqnarray}
A_{\alpha }^{\beta}(x,t)=A_{\alpha +}^{\beta}(x,t)+A_{\alpha -}^{\beta}(x,t).
\end{eqnarray}
Here corresponding $A_{\alpha}$'s are assumed to exist.
We would like to point out that the equation (\ref{trunKM}) is analogous to 
truncated Kramers-Moyal expansion. Two rather important special cases are,
 $0<\beta<1$ and $1< \beta < 2$. In the former case we get the operator
\begin{eqnarray}
{\cal{L}_\alpha^\beta}(x,t) &=&
I\!\!D_{x-}^{\beta}A_{\alpha -}^{\beta}(x,t)+ 
I\!\!D_{x+}^{\beta}A_{\alpha +}^{\beta}(x,t), 
\;\;\;\;\;\;\;\;\;\;\;\;\;\;\;\;\;\;\;\;\;\;\;\;\;\;0<\beta<1,\nonumber
\end{eqnarray}
and  in the latter case we get
\begin{eqnarray}
{\cal{L}_\alpha^\beta}(x,t)&=&-{\partial\over{\partial x}}
A_{\alpha}^1(x,t)\!+\!I\!\!D_{x-}^{\beta}A_{\alpha -}^{\beta}(x,t)\!+\! 
I\!\!D_{x+}^{\beta}A_{\alpha +}^{\beta}(x,t), \;\;\;\;1< \beta < 2.\nonumber
\end{eqnarray}
This operator can be identified as   generalizations of the Fokker-Planck 
operator (involving one space variable). 
It is clear that when $\alpha=1$ and $\beta=2$ we get back the
usual Fokker-Planck operator in~(\ref{eq:FPKop}).

Returning back to expression (\ref{mbeta}) it is clear that  the small time
behavior of different transitional moments decide the order of the 
derivative with respect to time (in order to demonstrate this point we
consider the example of a L\'evy process below).
On the other hand, small distance behavior of transitional probability or
the differentiability property of the initial probability density would
dictate the order of space derivative.
Depending on the actual values of 
$\alpha$ and $\beta$ as well as their interrelation the above local FFPK
equation will describe different processes. 

Equations which give rise to evolution-semigroup are of interest in physics.
The equation~(\ref{trunKM}) corresponds to a semigroup if $\alpha=1$. 
One can then write down a formal solution of the above equation in this case as
follows. In the time independent case we have
\begin{eqnarray}
W(x,t) = e^{{{\cal{L}}_{1}^\beta}(x) t}W(x,0)
\end{eqnarray} 
and  when the operator depends on time we have
\begin{eqnarray}
W(x,t) = \stackrel{\leftarrow}{T}e^{\int_0^t{{\cal{L}}_1^\beta}(x,t') dt'}W(x,0)
\end{eqnarray} 
where ${{\cal{L}}_1^\beta}$ is an operator in equation (\ref{trunKM}) and
 $\stackrel{\leftarrow}{T}$ is the time ordering operator.

\section{Examples}

\subsubsection{L\'evy processes}
The symmetric stable L\'evy process of index $\mu$, $0<\mu<2$, is defined by
\begin{eqnarray}
P(x,t) = \int_{-\infty}^{\infty} dk e^{-b|k|^\alpha t} e^{ikx}.
\end{eqnarray}
For this process all the moments, 
\begin{eqnarray}
M_\gamma^+(t) = \int_0^\infty dx x^\gamma \int_{-\infty}^{\infty} dk 
e^{-b|k|^\alpha t} e^{ikx},
\end{eqnarray}
of order  $\gamma \geq \mu$ diverge (see~\cite{5LM}). We have
\begin{eqnarray}
M_\gamma^+(\lambda t) &=& \int_0^\infty dx x^\gamma \int_{-\infty}^{\infty} dk e
^{-b|k|^\alpha \lambda t} e^{ikx} \\
&=& \lambda^{\gamma /\alpha} M_{\gamma}^+(t).
\end{eqnarray}
A similar relation holds for $M_\gamma^-$.
As a result the moments scale as
$M_\gamma(\lambda t) = \lambda^{\gamma /\mu} M_{\gamma}(t)$ and
the assumption in equation~(\ref{eq:intmnt}) is clearly not valid.
If we use our local FFPK then we get
\begin{eqnarray}
I\!\!D_t^{\gamma/ \mu}W(x,t)= 
I\!\!D_{x-}^{\gamma}[A_{\gamma/ \mu -}^{\gamma}(x,t) W(x,t)]+ 
I\!\!D_{x+}^{\gamma}[A_{\gamma/ \mu +}^{\gamma}(x,t) W(x,t)]. \label{FPL}
\end{eqnarray}
Since the process is symmetric the first derivative does not appear.
The order of the time derivative depends on that of space derivative but
it is always less than one. There is only one free parameter $\gamma$
which is restricted to be in the range $0 <\gamma <\mu$. 
In this case the value of $\gamma$ will be decided by the differentiability
class of the initial distribution function. 
When $\mu = 2$ we get a Gaussian process whose second moment is finite;
hence we can have
$\gamma = 2$ in the above equation and get 
back the usual Fokker-Planck equation. 
Equation (\ref{FPL}) forms one example where we get nontrivial 
(fractional) values for the 
orders of the
derivatives and usual 
derivation of FPK equation 
breaks down.

\subsubsection{A fractal time Gaussian process}
As our next example we consider the transition probability 
\begin{eqnarray}
P(x,t+\tau|x',t) &=& {1\over\sqrt{\pi\Delta{P}_C(t,\tau)}} 
 e^{-(x-x')^2\over{\Delta{P}_C(t,\tau)}}\label{eq:trP} \\
&=& \delta(x-x') \;\;\;\mbox{if} \;\;\;\Delta{P}_C(t,\tau) = 0,
\end{eqnarray}
where $\Delta{P}_C(t,\tau)=P_C(t+\tau)-P_C(t))$.
This transition probability describes a nonstationary process which corresponds 
to transitions occurring only at times which
lie on a fractal set. Such a transition probability can be used to model
phenomena such as diffusion in the 
presence of traps.
The second moment is given, from equation~(\ref{eq:moments}), by
\begin{eqnarray}
M_2(t,\tau) &=& {\Delta{P}_C(t,\tau)\over{2}} 
\simeq {1\over 2}{ I\!\!D^\alpha P_C(t) \over \Gamma(\alpha+1)} \tau^\alpha 
\nonumber\\
&=& {\tau^{\alpha}\over{2}} \chi_C(t)
\end{eqnarray}
This gives us the following local fractional Fokker-Planck equation 
(in this case an analog of a diffusion equation).
\begin{eqnarray}
I\!\!D_t^{\alpha}W(x,t) 
&=& {\Gamma(\alpha + 1)\over 4} \chi_C(t) {\partial^2\over{\partial x^2}}W(x,t)
\label{eq:ex} 
\end{eqnarray}
We note that even though the variable $t$ is taking all real positive values
the actual evolution takes place only for values of $t$ in the fractal set $C$.
The solution of equation~(\ref{eq:ex}) can easily be obtained as
\begin{eqnarray}
W(x,t) 
&=& P_{t-t_0}W(x,t_0)
\end{eqnarray}
where
\begin{eqnarray}
P_{t-t_0}
&=&  \lim_{N\rightarrow \infty} \prod_{i=0}^{N-1} \big[ 1 + {1\over 4} (t_{i+1}-t_{i})^\alpha F_C^i {\partial^2\over{\partial
x^2}}\big].
\end{eqnarray}
The above product converges because except for terms for which $F^i_C=1$  
(which are of order $N^{\alpha}$) all others take value 1.
It is clear that for $t_0 < t' < t$
\begin{eqnarray}
W(x,t) = P_{t-t'}P_{t'-t_0}W(x,t_0)
\end{eqnarray}
and $P_t$ gives rise to a semigroup evolution.
Using equation~(\ref{eq:PCt}) it can be 
easily seen that
\begin{eqnarray}
W(x,t) = e^{{P_C(t)\over 4}{\partial^2\over{\partial
x^2}}}W(x,t_0=0).
\end{eqnarray}
Now  choosing the initial distribution $W(x,0) = \delta(x)$ and using the
 Fourier representation
of delta function, we get the solution
\begin{eqnarray}
W(x,t) &=& {1\over\sqrt{\pi P_C(t) }}e^{ - x^2\over{P_C(t)}} \label{eq:sol}
\end{eqnarray}
Consistency of the equation~(\ref{eq:sol}) can easily be checked by 
directly substituting this in Chapman-Kolmogorov
equation~(\ref{CK}).
We note that this solution satisfies the bounds
\begin{eqnarray}
{1\over\sqrt{\pi bt^\alpha }}e^{ - x^2\over{bt^\alpha}} \leq 
W(x,t)  \leq
{1\over\sqrt{\pi at^\alpha }}e^{ - x^2\over{at^\alpha}}
\end{eqnarray}
for some $0<a<b<\infty$.
This is a model solution of a subdiffusive behavior. It is clear that
when $\alpha=1$ we get back the typical solution of the ordinary
diffusion equation, which is $(\pi t)^{-1/2}\exp(-x^2/t)$.

Anomalous diffusive behavior arises at numerous places in physics
and therefore obtaining such solutions from a kinetic equation is very
important. Specifically such transition probabilities 
(equation~(\ref{eq:trP})) may arise in
disordered systems when there is a diffusion in presence traps,
diffusion across cantori in phase space of the Hamiltonian systems 
or in phase space of weak chaotic Hamiltonian systems.

This work may have interesting consequences in other fields also. The LFDEs
that we have considered above suggest a generalization of Dynamical
systems to orders less than one. The solution that
we have found is also a solution 
of the Chapman-Kolmogorov equation defining Markov process. Therefore
our results may have relevance in probability theory. 
We would like to suggest that we have been able to incorporate
the phenomenon taking place in fractal time into an equation and find 
its exact solution. This marks a step towards development of calculus
for phenomena occurring in fractal space-time.

\chapter{Concluding Remarks}

\begin{itemize}
\item The thesis introduced and discussed a new notion 
called {\it local fractional 
derivative} of a function of one variable.
This is a modification of conventional fractional 
derivatives. A quantity, {\it critical order} of a function at a given point, 
has been defined as the maximum order
such that all the local fractional derivatives of orders below it exist. 

\item A fractional Taylor expansion has been constructed which contains these
LFDs as coefficients of fractional power law. This gives a geometrical 
interpretation to LFD.

\item We have established that (for sufficiently large $\lambda$ ) the critical order of the
Weierstrass' nowhere differentiable 
function is related to the box dimension of its graph. If  the dimension of
the graph of such a function is $1+\gamma$, the critical order is $1-\gamma$,
i.e., the function is differentiable upto order $1-\gamma$. When 
$\gamma$ approaches unity the function becomes more and more irregular and local fractional
differentiability is lost accordingly. Thus there is a direct quantitative connection between the 
dimension of the graph and the fractional differentiability property of the function.
A consequence of our result is that a classification of continuous paths 
(e.g. fractional Brownian paths) or
functions according to local fractional differentiability properties is also
a classification according to  dimensions (or H\"older exponents).

\item Also  the L\'evy index of a L\'evy flight on one dimensional 
lattice is identified as
the critical order of the characteristic function of the L\'evy walk. More generally,
the L\'evy index of the  distribution is identified as
the critical order of its characteristic function at the origin.

\item We have argued and demonstrated that LFDs are useful for studying isolated singularities and singularities masked by the stronger singularity (not just by
regular behavior). We have further shown that the pointwise
behavior of irregular, fractal or multifractal, functions can be studied
using LFDs.

\item We have also demonstrated that it is
possible to carry out the same theme even in the multidimensional case.
In particular, the H\"older exponents in any direction are related to
the critical order of the corresponding directional-LFD.
We feel that, whereas a one dimensional result is useful in studying
irregular signals, the results here may have utility in image processing
where one can characterize and classify singularities in the image data.
We note that it is possible to write a multivariable fractional Taylor
series which can be used for approximations and modeling of 
multivariable multiscaling functions. 
Further these considerations suggest a way for formulating fractional
differential geometry for fractal surfaces.

\item We have introduced new kind of differential equations called
local fractional differential equations. They are equations in terms of
LFDs. Our equations are fundamentally different
from any of the equations proposed previously,
since they involve LFDs. They are  local and
more natural generalization of ordinary differential equations.
LFDEs deserve a thorough study in their own right.

\item We have derived a generalization of FPK equation
which involves the local fractional derivatives. 
 Since the present analog of FPK equation is derived from 
first principles, we feel that our equations will have general
applicability in the field of physics. We expect them to be of value
in the studies of anomalous diffusion, disordered phenomena,
chaotic Hamiltonian systems, etc.

\end{itemize}

\end{document}